\documentclass[prl,twocolumn]{revtex4-2}
\usepackage{amsmath}
\usepackage{amssymb}
\usepackage[scr=boondox]{mathalpha}
\usepackage{graphicx}

\usepackage{tabularx}
\usepackage[table]{xcolor}
\usepackage{booktabs}

\usepackage{hyperref}

\usepackage{xcolor}
\newcommand{\R}[1]{#1}

\allowdisplaybreaks

\begin{document}

\title{Cumulative effects of laser-generated gravitational shock waves}

\author{Riccardo Falcone}
\affiliation{Department of Physics, University of Sapienza, Piazzale Aldo Moro 5, 00185 Rome, Italy}

\author{Claudio Conti}
\affiliation{Department of Physics, University of Sapienza, Piazzale Aldo Moro 5, 00185 Rome, Italy}

\begin{abstract}
The emission of light pulses is expected to generate gravitational waves, opening the possibility of controlling gravity in an Earthed laboratory. However, measuring the optically-driven spacetime deformations is challenging due to the inherently weak interaction. We explore the possibility to achieve a detectable gravitational effect from light emission by examining the cumulative effect of a sequence of laser-generated gravitational shock waves on a test particle. We derive an exact solution to the Einstein equations for cylindrically-shaped optical beams with constant energy density, imposing continuity condition for the metric and its first-order derivatives. Our analysis reveals that laser-induced gravitational fields cause a spatial shift in the test particle, which is measurable within current interferometric technology.
\end{abstract}

\maketitle

Gravitational Waves (GWs) have been directly detected for the first time in 2015 \cite{PhysRevLett.116.061102} by using highly sensitive interferometers capable of detecting changes in the lengths of their arms $L$ at a precision of $\Delta L / L \sim 10^{-22}$. Astronomical sources generate far greater power than anything achievable on Earth, but their distance and inaccessibility make them impractical for use. Generating metric perturbations in laboratory settings offers greater reproducibility and can be more precisely tailored to specific observational goals.

Proposals for generators of GWs in laboratory include the use of light beams \cite{PhysRev.37.602, Westervelt1965, Hegarty1969, PhysRevD.19.3582, Rätzel_2016, PhysRevD.95.084008, Schneiter_2018, PhysRevD.105.104052}, electromagnetic cavities \cite{Grishchuk:1973qz, 1975ZhETF..68.1569G, Leonid_P_Grishchuk_1977,  grishchuk2003electromagneticgeneratorsdetectorsgravitational, Spengler2022}, electromagnetic waves passing through a static homogeneous magnetic field \cite{Gertsenshtein1962, Zeldovich1983, PhysRevD.63.044014, Kolosnitsyn_2015} and lasers striking
a target \cite{10.1142/9789814374552_0292, 10.1063/1.4962520, Kadlecova2017}. Despite the enhanced results of laser-generated GWs compared to those from matter acceleration \cite{PhysRevD.105.104052, 10.1142/9789814374552_0292, Kadlecova2017}, detecting these perturbations in laboratory experiments remains challenging, as they are still $15$ orders of magnitude beyond the sensitivity of the most advanced detectors available today \cite{PhysRevD.105.104052}. In this work, we show that by exploiting the repeatability of laser pulse generation, it is possible to amplify the gravitational effects by several orders of magnitude. We investigate the perturbed motion of a free-falling particle influenced by a sequence of pulses, thereby allowing the gravitational effects of these pulses to accumulate over time and become compatible with the sensitivity of modern detectors. To accomplish this, we develop a theoretical framework describing the gravitational field generated by a sequence of laser pulses.

Historically, the gravitational effects of electromagnetic fields have been predominantly studied under the assumption of weak gravitational field and using the retarded integral approach \cite{PhysRev.37.602, Westervelt1965, Hegarty1969, PhysRevD.19.3582, Rätzel_2016, PhysRevD.95.084008, Schneiter_2018, PhysRevD.105.104052, Grishchuk:1973qz, 1975ZhETF..68.1569G, Leonid_P_Grishchuk_1977, grishchuk2003electromagneticgeneratorsdetectorsgravitational, Spengler2022, Gertsenshtein1962, Zeldovich1983, PhysRevD.63.044014, Kolosnitsyn_2015}. In their pioneering work, Tolman, Ehrenfest and Podolsky \cite{PhysRev.37.602} investigated the spacetime deformation induced by infinitesimally thin steady light beams and by passing light pulses, where diffraction effects were neglected. This topic was later revisited and expanded by Hegarty \cite{Hegarty1969}, who, following Westenvelt's earlier work \cite{Westervelt1965}, introduced the emitter and the absorber of the radiation as point sources. Hegarty further observed that the gravitational effects of the pulse originate solely from its emission and absorption, rather than from its propagation. More recently, Rätzel \textit{et al.}~\cite{Rätzel_2016} demonstrated that these gravitational effects are localized within spherical shells expanding at the speed of light, representing the spacetime imprints of the pulse emission and absorption events. To overcome the limitation of assuming zero radial extension, Lageyre \textit{et al.}~\cite{PhysRevD.105.104052} analyzed the spacetime deformation produced by a cylindrically shaped light pulse.

To go beyond the retarded integral approach, various studies have addressed the problem of solving the exact Einstein equations for a light pulse. A key advantage of the this method over the retarded integral approach is that it bypasses the need to model the exact process by which the laser pulse is emitted, allowing focus solely on the gravitational field near the pulse's trajectory \cite{Bonnor1969, PhysRevD.95.084008}. Exact solutions to the Einstein–Maxwell equations for plane electromagnetic waves are well-documented in the literature \cite{griffiths2016colliding} and have been expressed in both Brinkmann coordinates \cite{doi:10.1073/pnas.9.1.1, PhysRevLett.3.571, https://doi.org/10.1002/prop.201000088} and Rosen coordinates \cite{rosen1937plane, Bell1974, PhysRevD.89.104049}. In these cases, the source field generating the gravitational field is infinitely extended across spacetime and, therefore, cannot be used to model realistic light pulses. In contrast, Bonnor \cite{Bonnor1969} presented an exact solution for a null fluid traveling along an infinite straight path, accompanied by a GW propagating in the same direction as the pulse. The pulse is described by a stress-energy tensor of the form $T \propto (\partial_x^2 A + \partial_y^2 A) \partial_v \otimes \partial_v $, where $x$ and $y$ are the spacelike coordinates orthogonal to the direction of propagation, $u$ and $v$ are the co-propagating and counter-propagating null coordinates, and $A = A(u,v,x,y)$ is a $C^1$, piecewise $C^3$, function. The specific case of single photons was later examined by Aichelburg and Sexl \cite{Aichelburg1971}, who found that the associated gravitational field features a plane-fronted GW. Notably, this GW arises during the emission process rather than from the photon's propagation itself \cite{Bonnor2009}, in agreement with Hegarty \cite{Hegarty1969} and Rätzel \textit{et al.}~\cite{Rätzel_2016}.

Inspired by the findings by Hegarty \cite{Hegarty1969}, Bonnor \cite{Bonnor2009}, Rätzel \textit{et al.}~\cite{Rätzel_2016}, we investigate the propagation of a pulse and an accompanying GW, which locally and at large distances from the source appears to share the same plane wavefront as the pulse. Following Ref.\ \cite{PhysRevD.105.104052}, the pulse is modeled as a cylindrical beam with constant energy density. We solve the Einstein equations for this configuration, ensuring continuity of the metric and its first-order derivatives \footnote{Discontinuities in the second-order derivatives are expected, as the Einstein equation is a second-order partial differential equation, and the stress-energy tensor is discontinuous at the border of the pulse.}. This continuity condition, known as the Lichnerowicz condition \cite{lichnerowicz1955théories}, guarantees that the curvature tensor is piecewise $C^0$ and nowhere suffers anything worse than a straight shock discontinuity. Due to discontinuities of the stress-energy tensor with respect to the $u$ coordinate, Bonnor's solution \cite{Bonnor1969} cannot be applied to the entire spacetime. To address this, we partition the spacetime into separate patches: In the region containing the pulse and the GW, we employ Bonnor's solution for a steady uniform beam with a circular cross-section. We then use the Brinkmann-Rosen coordinate transformation to define alternative coordinate charts that can be smoothly extended into the Minkowski exterior region. The resulting nontrivial description of the spacetime cannot be captured by the single coordinate system used in the retarded integral approach. After deriving the spacetime solution for a single pulse, we extend our analysis to a sequence of pulses with a constant repetition rate. The metric solutions for individual pulses are patched together and the trajectory of a particle 
under the influence of this sequence of pulses is then computed.

\begin{figure}[h]
\includegraphics[]{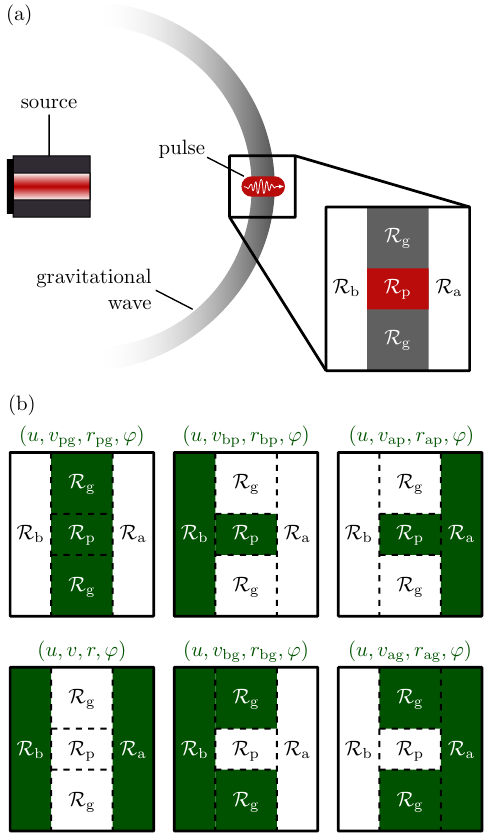}
\caption{We consider an electromagnetic pulse propagating alongside a gravitational wave (GW), both originating from a distant source. Locally, the spacetime is divided into four distinct regions: $\mathcal{R}_\text{p}$, $\mathcal{R}_\text{g}$, $\mathcal{R}_\text{a}$ and $\mathcal{R}_\text{b}$ [panel (a)]. The regions $\mathcal{R}_\text{p}$ and $\mathcal{R}_\text{g}$ correspond to the pulse and the GW, respectively, while $\mathcal{R}_\text{a}$ and $\mathcal{R}_\text{b}$ represent the Minkowski spacetime before and after the passage of the pulse and the GW. Panel (b) shows the various coordinate systems used to describe the spacetime, each covering different combinations of regions. The coordinates $(u, v_{\text{p}\text{g}}, r_{\text{p}\text{g}}, \varphi)$ are used for the interior regions $\mathcal{R}_\text{p}$ and $\mathcal{R}_\text{g}$, while $(u, v, r, \varphi)$ are the standard Minkowski coordinates for $\mathcal{R}_\text{a} \cup \mathcal{R}_\text{b}$. Additionally, for each $\xi \R{\in} \{ \text{a}, \text{b} \}$ and $\eta \R{\in} \{ \text{p}, \text{g} \}$, the coordinate system $(u, v_{\xi \eta}, r_{\xi \eta}, \varphi)$ spans the region $\mathcal{R}_\xi \cup \mathcal{R}_\eta$. }\label{Single_pulse_figure}
\end{figure}

\textit{Spacetime of a single pulse}---We start by considering the propagation of a single pulse whose source lies far outside the region of interest. When the pulse is emitted, it also generates a nonzero GW, which undergoes diffraction, while the pulse itself remains confined to a finite spatial region over the observed timeframe. Because the source is very distant, the local spacetime appears indistinguishable from that of a pulse originating from past null infinity. In what follows, we offer a general description of this spacetime, with details provided in the Supplemental Material \cite{Supplemental}.

We note by $\mathcal{R}_\text{p}$, $\mathcal{R}_\text{g}$ and $\mathcal{R}_\text{M}$ the spacetime regions describing the pulse, the GW and the exterior, respectively [Fig.\ \ref{Single_pulse_figure} panel (a)]. The region $\mathcal{R}_\text{M}$ is further divided into $\mathcal{R}_\text{a}$ and $\mathcal{R}_\text{b}$, corresponding to the areas before and after the passage of the wave. As no gravitational source acts on $\mathcal{R}_\text{M}$, the exterior region can be described by a Minkowski coordinate system $(u, v, r, \varphi)$, with metric
\begin{equation}\label{metric_Minkowski}
\left.  ds^2 \right|_{\mathcal{R}_\text{M}} = -2dudv + dr^2 +r^2 d\varphi^2.
\end{equation}
Here, $u$ and $v$ are the light-cone coordinates defined via $u=(ct-z)/\sqrt{2}$ and $v=(ct+z)/\sqrt{2}$, with $c$ as the speed of light, $t$ as the time coordinate and $z$ as the direction of the propagation; $r$ and $ \varphi$ are the radial coordinates orthogonal to $z$.

At a large distance from the source, the pulse and the GW appear with a plane wavefront traveling in the same direction. This suggest the use of the same null coordinate $u$ for all regions ($\mathcal{R}_\text{p}$, $\mathcal{R}_\text{g}$ and $\mathcal{R}_\text{M}$). Likewise, by assuming axial symmetry around the central axis of the beam, we consider a global azimuthal coordinate $\varphi$. Since the wavefront is planar, we define two hyperplanes of constant $u$, labeled $\bar{u}_\text{a}$ and $\bar{u}_\text{a}$, which mark the boundaries of the exterior regions $\mathcal{R}_\text{a}$ and $\mathcal{R}_\text{b}$. Specifically, $\mathcal{R}_\text{a}$ corresponds to $u<\bar{u}_\text{a}$, whereas $\mathcal{R}_\text{b}$ corresponds to $u>\bar{u}_\text{b}$.

We model the pulse as a cylinder of light with constant energy density, traveling in the $z$ direction with velocity $c$. This setup can approximate a laser beam with circular polarization and a flat profile (e.g., flattened Gaussian pulsed beam \cite{GORI1994335}). As a consequence of this approximation, the spacetime within the pulse exhibits cylindrical symmetry. Such a symmetry suggests a preferred choice of coordinate system $(u, v_{\text{p}\text{g}}, r_{\text{p}\text{g}}, \varphi)$ for $\mathcal{R}_\text{p} \cup \mathcal{R}_\text{g}$, in which $\partial_{z_{\text{p}\text{g}}} = (\partial_{v_{\text{p}\text{g}}} - \partial_u)/\sqrt{2}$ and $\partial_\varphi$ are Killing vector fields.

The metric associated to $(u, v_{\text{p}\text{g}}, r_{\text{p}\text{g}}, \varphi)$ is
\begin{equation}\label{g_vr_pg}
\left.  ds^2 \right|_{\mathcal{R}_\text{p} \cup \mathcal{R}_\text{g}} = - \Phi(r_{\text{p}\text{g}}) du^2 -2dudv_{\text{p}\text{g}} + dr_{\text{p}\text{g}}^2 +r_{\text{p}\text{g}}^2 d\varphi^2,
\end{equation}
with
\begin{equation}\label{Phi}
\Phi(r_{\text{p}\text{g}}) = \begin{cases}
\epsilon (r_{\text{p}\text{g}}/\bar{r}_\text{p})^2 & \text{if } r_{\text{p}\text{g}} \leq \bar{r}_\text{p},  \\
\epsilon \left[1 + 2\log(r_{\text{p}\text{g}}/\bar{r}_\text{p})  \right] & \text{if } r_{\text{p}\text{g}} > \bar{r}_\text{p}.
\end{cases}
\end{equation}
Here, $\bar{r}_\text{p}$ represents the radius of the pulse and $\epsilon = \kappa I \bar{r}_\text{p}^2 /c$ is the coupling between the pulse and the gravitational field, with $\kappa$ as the Einstein gravitational constant and $I$ as the intensity of the electromagnetic field. Equation (\ref{g_vr_pg}) provides the solution to the Einstein equation with $T = \theta(\bar{r}_\text{p} - r_{\text{p}\text{g}})  ( 2 I /c) \partial_{v_{\text{p}\text{g}}} \otimes \partial_{v_{\text{p}\text{g}}}$ as the stress-energy tensor and with $\partial_{z_{\text{p}\text{g}}} = (\partial_{v_{\text{p}\text{g}}} - \partial_u )/\sqrt{2}$ and $\partial_\varphi$ as the Killing fields associated with the cylindrical symmetry. This solution is equivalent to the one presented by Bonnor \cite{Bonnor1969} for the case of a steady uniform beam of circular cross section.

To connect the coordinate systems $(u, v, r, \varphi)$ and $(u, v_{\text{p}\text{g}}, r_{\text{p}\text{g}}, \varphi)$  across the hypersurfaces $u = \bar{u}_\text{a}$ and $u = \bar{u}_\text{b}$, we introduce additional charts that cover distinct subregions of the full spacetime, thereby creating a complete atlas [Fig.\ \ref{Single_pulse_figure} panel (b)]. By imposing the continuity of the metric components $g_{\mu\nu}$ (up to their first-order derivatives) across $u = \bar{u}_\text{a}$ and $u = \bar{u}_\text{b}$, we obtain four distinct coordinate systems, denoted as $(u, v_{\xi \eta}, r_{\xi \eta}, \varphi)$, with $\xi \R{\in} \{ \text{a}, \text{b} \}$ and $\eta \R{\in} \{ \text{p}, \text{g} \}$, each covering the region $\mathcal{R}_\xi \cup \mathcal{R}_\eta$. In $\mathcal{R}_\eta$, the coordinates $(u, v_{\xi \eta}, r_{\xi \eta}, \varphi)$ are defined by the transformation
\begin{align}\label{vr_pg_vr_xieta}
& v_{\text{p}\text{g}} = v_{\xi \eta} + V_\eta \left(u-\bar{u}_\xi ,r_{\xi\eta} \right), && r_{\text{p}\text{g}} = R_\eta \left(u-\bar{u}_\xi ,r_{\xi\eta} \right),
\end{align}
with
\begin{align}
& \begin{aligned}[t] & V_{\text{p}}(u,r) = \frac{R_{\text{p}}(u,r) \dot{R}_{\text{p}}(u,r)}{2}, && R_{\text{p}}(u,r) = \cos \left( \sqrt{\epsilon} \frac{u}{\bar{r}_{\text{p}}} \right) r, \\ \end{aligned} \nonumber \\
& V_{\text{g}}(u, r) =  R_{\text{g}}(u,r) \dot{R}_{\text{g}}(u,r) + \frac{\epsilon}{2}  \left[1-2 \log \left(\frac{r}{\bar{r }_{\text{p}}}\right)\right] u, \nonumber  \\
& R_{\text{g}}(u, r) = \exp \left\lbrace-\left[\text{erf}^{-1}\left(\sqrt{\frac{2 \epsilon }{\pi }} \frac{u }{r }\right)\right]^2\right\rbrace r, \nonumber  \\
& \dot{R}_\eta(u,r) = \partial_u R_\eta(u,r),
\end{align}
and $\text{erf}^{-1}(\cdot)$ as the inverse error function \footnote{The domain of inverse error function is $(-1,1)$. In this paper, we assume $\epsilon \ll \bar{r}_\text{p}^2/(\bar{u}_\text{b}-\bar{u}_\text{a})^2$, which ensures that the transformation (\ref{vr_pg_vr_xieta}) is well-defined throughout the entire region $\mathcal{R}_\text{g}$. If this condition is not satisfied, it becomes necessary to define subregions $\mathcal{D}_\xi \subset \mathcal{R}_\text{g}$ as $| u - \bar{u}_\xi| / r_{\xi\text{g}} < \sqrt{\pi/2 \epsilon}$ and assume that the coordinate systems $(u, v_{\xi \text{g}}, r_{\xi \text{g}}, \varphi)$ cover the subregions $\mathcal{R}_\xi \cup \mathcal{D}_\xi \subset \mathcal{R}_\xi \cup \mathcal{R}_\text{g}$ instead of $\mathcal{R}_\xi \cup \mathcal{R}_\text{g}$. This modification does not affect the results of this paper, as the coordinate systems $(u, v_{\text{p} \text{g}}, r_{\text{p} \text{g}}, \varphi)$ and $(u, v_{\xi \eta}, r_{\xi \eta}, \varphi)$ still provide a complete atlas covering both the interior and exterior regions.}. The resulting metric is
\begin{align}
\left.  ds^2 \right|_{\mathcal{R}_\eta}  = & - 2 du dv_{\xi\eta} + \left[ R'_\eta(u- \bar{u}_\xi, r_{\xi\eta}) \right]^2 dr_{\xi\eta}^2 \nonumber \\
& +  R_\eta^2 (u- \bar{u}_\xi, r_{\xi\eta}) d\varphi^2,
\end{align}
with $R'_\eta(u,r) = \partial_r R_\eta(u,r)$. \R{In this coordinate system, the metric explicitly exhibits asymptotic flatness within $\mathcal{R}_\text{g}$, as $R_{\text{g}}(u, r)\sim r$ for $r \to \infty$.} In the exterior region $\mathcal{R}_\xi$, the coordinates $(u, v_{\xi \eta}, r_{\xi \eta}, \varphi)$ are equivalent to the Minkowski ones: $ v = v_{\xi \eta}$ and $ r = r_{\xi\eta}$.

\textit{Particle trajectories}---In the exterior region $\mathcal{R}_\text{M}$, particles travel with constant velocity due to the flatness of the metric. However, their trajectories are perturbed upon interacting with the GW. To analyze this effect, we solve the geodesic equation within the region $\mathcal{R}_\text{g}$ by using Eq.\ (\ref{g_vr_pg}) and match these solutions with the straight-line geodesics in $\mathcal{R}_\text{M}$ at the boundaries $u = \bar{u}_\xi$ through the coordinate transformations $ (u, v_{\xi\text{g}}, r_{\xi\text{g}}, \varphi) \mapsto (u, v_{\text{p}\text{g}}, r_{\text{p}\text{g}}, \varphi)$.

We consider a particle that traverses $\mathcal{R}_\text{a}$, $\mathcal{R}_\text{g}$ and $\mathcal{R}_\text{b}$ while remaining confined to the semiplane defined by $\varphi = 0$.  Under these assumptions, the general solution to the geodesic equation is
\begin{align}\label{gamma_eta}
& (u, v_{\text{p}\text{g}}, r_{\text{p}\text{g}}, \varphi) = \left( c_{\text{g},1} s , c_{\text{g},2} s + c_{\text{g},3} \right. \nonumber \\
& \left. + V_{\text{g}}\left(c_{\text{g},1} s + c_{\text{g},4}, c_{\text{g},5} \right), R_{\text{g}}\left(c_{\text{g},1} s + c_{\text{g},4},c_{\text{g},5}\right) , 0 \right)
\end{align}
as detailed in the Supplemental Material \cite{Supplemental}. The parameters $c_{\text{g},1}, \dots, c_{\text{g},5}$ are determined by the boundary conditions. In addition to Eq.\ (\ref{gamma_eta}), we also consider the geodesics in the exterior regions $\mathcal{R}_\text{a}$ and $\mathcal{R}_\text{b}$, which take the form
\begin{equation}\label{gamma_xi}
(u, v_{\xi\text{g}}, r_{\xi\text{g}}, \varphi) = \left( c_{\xi,1} s , c_{\xi,2} s + c_{\xi,3},  c_{\xi,4} s + c_{\xi,5} , 0 \right).
\end{equation}
The matching conditions across the boundary between $\mathcal{R}_\xi$ and $\mathcal{R}_\text{g}$ are
\begin{align} \label{c_xi_c_g}
& \begin{aligned} & c_{\xi ,1} = c_{\text{g},1},  \nonumber \\
& c_{\xi ,2} = c_{\text{g},2} + \frac{c_{\text{g},1}}{2}\left[ \dot{R}_{\text{g}} \left(\bar{u}_{\xi }+c_{\text{g},4},c_{\text{g},5}\right) \right]^2,  \end{aligned} \nonumber \\
& \begin{aligned} c_{\xi ,3} = \, & c_{\text{g},3} + V_{\text{g}}\left(\bar{u}_{\xi }+c_{\text{g},4},c_{\text{g},5}\right) \nonumber \\
& -\frac{1}{2} \left[ \dot{R}_{\text{g}} \left(\bar{u}_{\xi }+c_{\text{g},4},c_{\text{g},5}\right) \right]^2 \bar{u}_{\xi }, \end{aligned} \nonumber \\
& c_{\xi ,4} = c_{\text{g},1} \dot{R}_{\text{g}}\left(\bar{u}_{\xi }+c_{\text{g},4},c_{\text{g},5}\right), \nonumber \\
& c_{\xi ,5} = R_{\text{g}}\left(\bar{u}_{\xi }+c_{\text{g},4},c_{\text{g},5}\right)-\dot{R}_{\text{g}}\left(\bar{u}_{\xi }+c_{\text{g},4},c_{\text{g},5}\right)\bar{u}_{\xi } .
\end{align}

We assume that the particle enters the region $\mathcal{R}_\text{g}$ at the event $ (u,v,r,\varphi) = (\bar{u}_\text{a}, \bar{v}_\text{a}, \bar{r}_\text{a}, 0)$ and will exit $\mathcal{R}_\text{g}$ at $ (u,v,r,\varphi) = (\bar{u}_\text{b}, \bar{v}_\text{b}, \bar{r}_\text{b}, 0)$. The initial and the final four-velocities are denoted as $\left. U \right|_{u=\bar{u}_\text{a}} = U_\text{a}^u \partial_u + U_\text{a}^v \partial_v + U_\text{a}^r \partial_r$ and $\left. U \right|_{u=\bar{u}_\text{b}} = U_\text{b}^u \partial_u + U_\text{b}^v \partial_v + U_\text{b}^r \partial_r$, respectively. The explicit mapping of the initial data $\bar{v}_\text{a}, \bar{r}_\text{a}, U_\text{a}^u, U_\text{a}^v, U_\text{a}^r$ to the final quantities $ \bar{v}_\text{b}, \bar{r}_\text{b}, U_\text{b}^u, U_\text{b}^v, U_\text{b}^r$ can be derived by using Eq.\ (\ref{c_xi_c_g}) with $\xi=\text{b}$ and the inverse of Eq.\ (\ref{c_xi_c_g}) with $\xi=\text{a}$ 
\cite{Supplemental}.

\R{In the limit $\bar{r}_\text{a} \to \infty$, the map $(\bar{v}_\text{a}, \bar{r}_\text{a}, U_\text{a}^u, U_\text{a}^v, U_\text{a}^r) \mapsto (\bar{v}_\text{b}, \bar{r}_\text{b}, U_\text{b}^u, U_\text{b}^v, U_\text{b}^r)$ leads to $U_{\text{b}}^u \approx U_{\text{a}}^u$, $ U_{\text{b}}^v \approx U_{\text{a}}^v $ and $ U_{\text{b}}^r \approx U_{\text{a}}^r$ \cite{Supplemental}, indicating that the test particle does not accelerate in the asymptotic region, even when transitioning from $\mathcal{R}_\text{a}$ to $\mathcal{R}_\text{b}$. As a result, the coordinate system $(u, v, r, \varphi)$ can be interpreted as a laboratory frame. The map $(\bar{v}_\text{a}, \bar{r}_\text{a}, U_\text{a}^u, U_\text{a}^v, U_\text{a}^r) \mapsto (\bar{v}_\text{b}, \bar{r}_\text{b}, U_\text{b}^u, U_\text{b}^v, U_\text{b}^r)$ thus describes the motion of a test particle relative to laboratory particles situated far from the pulse.}

We focus on the specific case where $\epsilon \ll 1$, corresponding to weak gravitational effects. Also, we assume that the particle is initially at rest in $\mathcal{R}_\text{a}$, meaning that $U_\text{a}^u = U_\text{a}^v = 1/\sqrt{2}$ and $U_\text{a}^r=0$. We find that, after interacting with the GW, the particle 
is described by
\begin{align}
& \bar{v}_\text{b} \approx  \bar{v}_{\text{a}} +  \bar{u}_{\text{b}} - \bar{u}_{\text{a}}  - \epsilon \left[ \frac{1}{2} + \log \left( \frac{\bar{r}_{\text{a}}}{\bar{r}_{\text{p}}}\right) \right] \left( \bar{u}_{\text{b}} - \bar{u}_{\text{a}} \right)  ,\nonumber \\
& \begin{aligned} & \bar{r}_\text{b} \approx \bar{r}_{\text{a}}  - \frac{\epsilon}{2} \frac{\left( \bar{u}_{\text{b}} - \bar{u}_{\text{a}} \right) ^2}{\bar{r}_{\text{a}}} ,  & U_{\text{b}}^u = \frac{1}{\sqrt{2}}, \end{aligned}  \nonumber \\
& \begin{aligned} & U_\text{b}^v \approx \frac{1}{\sqrt{2}} +\frac{\epsilon^2}{2 \sqrt{2}} \left(\frac{ \bar{u}_{\text{b}}-\bar{u}_{\text{a}}}{\bar{r}_{\text{a}}}\right)^2 , & U_\text{b}^r \approx -\frac{\epsilon}{\sqrt{2}}  \frac{ \bar{u}_{\text{b}} - \bar{u}_{\text{a}}}{\bar{r}_{\text{a}}} . \end{aligned}
\end{align}

\textit{Sequence of pulses}---In typical laboratory settings, the parameter $\epsilon$ is extremely small, making it difficult to detect the perturbation in the particle trajectory. Nevertheless, the effect accumulates over the passage of multiple pulses, ultimately becoming measurable.

\begin{figure}[h]
\includegraphics[]{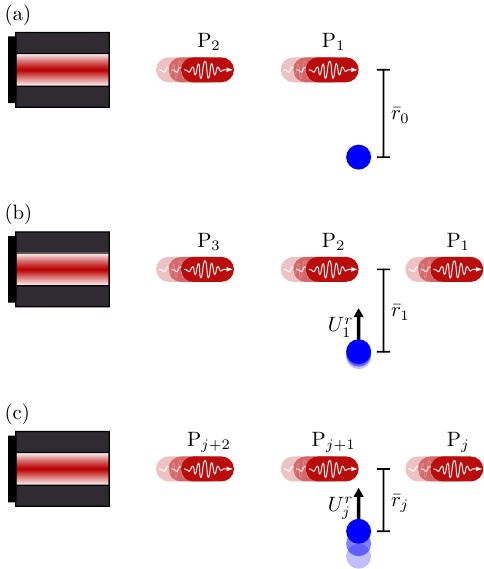}
\caption{We consider a particle initially at rest with a radial coordinate $r = \bar{r}_0$ [panel (a)]. As a consequence of the passage of the pulse $\text{P}_1$, the particle accelerates towards the pulse, acquiring a nonzero radial component of its four-velocity $U_1^r$ and experiencing a radial displacement $\bar{r}_1 - \bar{r}_0 < 0$ [panel (b)]. The effect accumulates with the sequential passage of additional pulses $\text{P}_2, \dots, \text{P}_j$, leading to a measurable radial displacement $\bar{r}_j - \bar{r}_0$ at $t=\bar{t}_j$ [panel (c)]. }\label{Sequence_of_pulses_figure}
\end{figure}

Here, we examine the spacetime of a sequence of pulses and its effect on the trajectory of a particle initially at rest [Fig.\ \ref{Sequence_of_pulses_figure}]. The $j$-th pulse $\text{P}_j$ and its associated GW are confined to the region $\bar{u}_j < u < \bar{u}_j + c \tau/\sqrt{2}$, where $\tau$ is the duration of the pulse. We assume a constant repetition rate $f_\text{rep}$, such that $\bar{u}_j = \bar{u}_{j-1} + c/\sqrt{2}f_\text{rep} $ for all $j$. The interior and exterior regions of each pulse are modeled by using the single-pulse spacetime described earlier, with successive pulses connected by a shared in-between Minkowski spacetime.

The geodesics are locally described by Eq.\ (\ref{gamma_eta}) within the interior regions and by Eq.\ (\ref{gamma_xi}) in the exterior regions. By recursively applying Eq.\ (\ref{c_xi_c_g}) and its inverse at each boundary $ u = \bar{u}_j$ and $u = \bar{u}_j + c \tau/\sqrt{2}$, we derive the trajectory of a particle influenced by the GWs. The explicit recursive equation is detailed in the Supplemental Material \cite{Supplemental}. Here, we focus on the case where $\epsilon \ll 1$, which allows for an analytic solution to the equation.

We assume that the particle begins interacting with the GW of the first pulse at $t = 0$. At this initial moment, its position is specified in Minkowski coordinates as $z=\bar{z}_0$ and $r=\bar{r}_0$. The particle is assumed to be initially at rest, so its four-velocity at $t = 0$ is $\left. U \right|_{t=0} = c^{-1}\partial_t $. The time and position of the particle when the $(j+1)$-th GW arrives are denoted as $\bar{t}_j$, $\bar{z}_j$ and $\bar{r}_j$. At that moment, the four-velocity is given by $\left. U \right|_{t=\bar{t}_j} = U_j^t \partial_t + U_j^z \partial_z + U_j^r \partial_r $. As a result of our analysis, we find:
\begin{align}\label{tzu_j}
& \bar{t}_j \approx j \frac{c}{f_{\text{rep}}}  - \frac{ \epsilon  j }{2}   \left[\frac{1}{2} + \log \left(\frac{\bar{r}_0}{\bar{r}_{\text{p}}}\right)\right] c \tau , \nonumber \\
& \bar{z}_j \approx  \bar{z}_0 - \frac{ \epsilon  j }{2} \left[\frac{1}{2} + \log \left(\frac{\bar{r}_0}{\bar{r}_{\text{p}}}\right)\right] c \tau , \nonumber \\
& \bar{r}_j \approx \bar{r}_0 -  \frac{\epsilon  j \left(j+1-\tau  f_{\text{rep}}\right)}{4}  \frac{c^2 \tau}{\bar{r}_0 f_{\text{rep}}}
\end{align}
and
\begin{align}\label{U_tzu_j}
& U_j^t \approx 1 + \frac{ \epsilon ^2 j^2 }{8}  \frac{c^2 \tau^2}{ \bar{r}_0^2}, && U_j^z \approx \frac{ \epsilon ^2 j^2 }{8}  \frac{c^2 \tau^2}{ \bar{r}_0^2}, && U_j^r \approx -\frac{\epsilon  j }{2 } \frac{c \tau }{\bar{r}_0}.
\end{align}

A large number of pulses ($j \gg 1$) can partly compensate for the weak gravitational interaction ($\epsilon \ll 1$). In the regime where $j \gg 1$ and $\epsilon j \ll 1$, the perturbation to the particle radial position becomes the dominant effect in Eqs.\ (\ref{tzu_j}) and (\ref{U_tzu_j}), as it scales as $|\bar{r}_j - \bar{r}_0| \sim \epsilon j^2$:
\begin{equation}\label{r_j_r_0}
\bar{r}_j - \bar{r}_0 \approx -  \frac{\epsilon  j^2}{4}  \frac{c^2 \tau}{\bar{r}_0 f_{\text{rep}}}.
\end{equation}
Figure \ref{Sequence_of_pulses_figure} illustrates the cumulative radial displacement of the particle caused by the sequence of pulses.

\textit{Concluding remarks}---To simulate an experimentally achievable scenario, we consider an ultra-intense pulsed laser, with the following parameters: $I = 10^{22} \text{ W}/\text{cm}^2$, $\bar{r}_{\text{p}} = 5 \text{ $\mu $m}$, $c \tau = 20 \text{ $\mu $m}$, $f_{\text{rep}} = 10^3 \text{ Hz}$. For the initial conditions, we assume 
$\bar{r}_0 = 500 \text{ $\mu $m}$. After $36$ million pulses (equivalent to $10$ hours of operation), the radial position will show a perturbation of approximately $\bar{r}_{36 \times 10^6}- \bar{r}_0 \approx -7 \times 10^{-18} \text{ m}$.

\begin{figure}[h]
\includegraphics[width=\columnwidth]{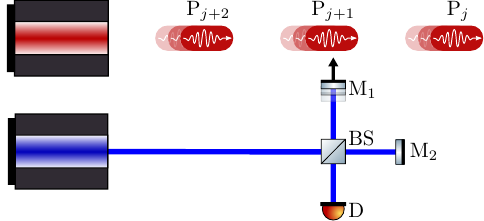}
\caption{The cumulative effect of laser-generated gravitational shock waves can be detected by using an interferometric setup consisting of a beam splitter (BS), two mirrors ($\text{M}_1$ and $\text{M}_2$), and a detector (D). Mirror $\text{M}_1$ is designed to move freely along the direction perpendicular to the propagation of the pulses $\text{P}_1, \dots, \text{P}_j, \dots$ which generate the gravitational field. The size of $\text{M}_1$ must be smaller than the size of each pulse $\text{P}_j$ (approximately on the order of a micron). The cumulative effects of the gravitational shock waves on $\text{M}_1$ produces a displacement that can be measured at D through interference.}\label{Measurement_figure}
\end{figure}

The LIGO interferometer detected gravitational strains on the order of $10^{-22}$ by using $4 \text{km}$-long arms, corresponding to a path length difference of approximately $\Delta L \sim 10^{-19} \text{m}$ ($10^{-4}$ times the width of a proton) \cite{PhysRevLett.116.061102}. Based on our calculations, the displacement of a particle influenced by $36$ million pulses falls within this precision range. \R{Additionally, the repetition rate $f_{\text{rep}} = 10^3 \text{ Hz}$ lies within the sensitivity bandwidth of LIGO, between 10 Hz and 1000 Hz.} This finding suggests the feasibility of using an interferometric setup, as depicted in Fig.\ \ref{Measurement_figure}, where one of the mirrors acts as the free-falling particle. Our analysis demonstrates that such a setup allows the measurement of the cumulative effect of laser-generated gravitational shock waves using existing technology.

We remark that according to Eq.\ (\ref{r_j_r_0}), the displacement of the test particle scales quadratically with the number of pulses $j$. Hence, increasing the observation time significantly amplifies the effect. For examples, a 
picometer displacement can be achieved within one year of accumulation. Also, since the gravitational shocks are generated by a sequence of laser pulses, various techniques can be employed to isolate the relevant information from background noise. The detected signal will exhibit a spectral content centered on the repetition rate of the laser, allowing methods such as heterodyne mixing and lock-in detection to enhance the signal-to-noise ratio by several orders of magnitude. In addition, one can superimpose slow-timescale modulations in the amplitude or the polarization of the laser pulses to discern spectral features that further improve measurement sensitivity.

These findings have important implications also for quantum gravity, particularly in the context of measuring gravitational effects arising from sources in quantum mechanical superposition \cite{Pikovski2012, Schmöle_2016, PhysRevD.98.126009, PhysRevLett.119.240402}. Light pulses may offer a promising alternative to large masses as generators of gravitational fields in quantum superpositions, especially given the current capability to produce cat states with laser beams \cite{Lewenstein2021}. Additionally, the gravitational effects of these systems can only be understood in the context of general relativity, thus offering a way to probe quantum degrees of freedom of gravity as a field (see Ref.\ \cite{chen2024quantumeffectsgravitynewton} and references therein).

\textit{Acknowledgment}---We acknowledge financial support by the HORIZON-EIC-2022-PATHFINDERCHALLENGES-01 HEISINGBERG project 101114978.

\bibliographystyle{apsrev4-2}
\bibliography{bibliography}

\begin{thebibliography}{42}%
\makeatletter
\providecommand \@ifxundefined [1]{%
 \@ifx{#1\undefined}
}%
\providecommand \@ifnum [1]{%
 \ifnum #1\expandafter \@firstoftwo
 \else \expandafter \@secondoftwo
 \fi
}%
\providecommand \@ifx [1]{%
 \ifx #1\expandafter \@firstoftwo
 \else \expandafter \@secondoftwo
 \fi
}%
\providecommand \natexlab [1]{#1}%
\providecommand \enquote  [1]{``#1''}%
\providecommand \bibnamefont  [1]{#1}%
\providecommand \bibfnamefont [1]{#1}%
\providecommand \citenamefont [1]{#1}%
\providecommand \href@noop [0]{\@secondoftwo}%
\providecommand \href [0]{\begingroup \@sanitize@url \@href}%
\providecommand \@href[1]{\@@startlink{#1}\@@href}%
\providecommand \@@href[1]{\endgroup#1\@@endlink}%
\providecommand \@sanitize@url [0]{\catcode `\\12\catcode `\$12\catcode
  `\&12\catcode `\#12\catcode `\^12\catcode `\_12\catcode `\%12\relax}%
\providecommand \@@startlink[1]{}%
\providecommand \@@endlink[0]{}%
\providecommand \url  [0]{\begingroup\@sanitize@url \@url }%
\providecommand \@url [1]{\endgroup\@href {#1}{\urlprefix }}%
\providecommand \urlprefix  [0]{URL }%
\providecommand \Eprint [0]{\href }%
\providecommand \doibase [0]{https://doi.org/}%
\providecommand \selectlanguage [0]{\@gobble}%
\providecommand \bibinfo  [0]{\@secondoftwo}%
\providecommand \bibfield  [0]{\@secondoftwo}%
\providecommand \translation [1]{[#1]}%
\providecommand \BibitemOpen [0]{}%
\providecommand \bibitemStop [0]{}%
\providecommand \bibitemNoStop [0]{.\EOS\space}%
\providecommand \EOS [0]{\spacefactor3000\relax}%
\providecommand \BibitemShut  [1]{\csname bibitem#1\endcsname}%
\let\auto@bib@innerbib\@empty
\bibitem [{\citenamefont {Abbott}\ \emph {et~al.}(2016)\citenamefont {Abbott}
  \emph {et~al.}}]{PhysRevLett.116.061102}%
  \BibitemOpen
  \bibfield  {author} {\bibinfo {author} {\bibfnamefont {B.~P.}\ \bibnamefont
  {Abbott}} \emph {et~al.} (\bibinfo {collaboration} {LIGO Scientific
  Collaboration and Virgo Collaboration}),\ }\href
  {https://doi.org/10.1103/PhysRevLett.116.061102} {\bibfield  {journal}
  {\bibinfo  {journal} {Phys. Rev. Lett.}\ }\textbf {\bibinfo {volume} {116}},\
  \bibinfo {pages} {061102} (\bibinfo {year} {2016})}\BibitemShut {NoStop}%
\bibitem [{\citenamefont {Tolman}\ \emph {et~al.}(1931)\citenamefont {Tolman},
  \citenamefont {Ehrenfest},\ and\ \citenamefont {Podolsky}}]{PhysRev.37.602}%
  \BibitemOpen
  \bibfield  {author} {\bibinfo {author} {\bibfnamefont {R.~C.}\ \bibnamefont
  {Tolman}}, \bibinfo {author} {\bibfnamefont {P.}~\bibnamefont {Ehrenfest}},\
  and\ \bibinfo {author} {\bibfnamefont {B.}~\bibnamefont {Podolsky}},\ }\href
  {https://doi.org/10.1103/PhysRev.37.602} {\bibfield  {journal} {\bibinfo
  {journal} {Phys. Rev.}\ }\textbf {\bibinfo {volume} {37}},\ \bibinfo {pages}
  {602} (\bibinfo {year} {1931})}\BibitemShut {NoStop}%
\bibitem [{\citenamefont {Westervelt}(1965)}]{Westervelt1965}%
  \BibitemOpen
  \bibfield  {author} {\bibinfo {author} {\bibfnamefont {P.}~\bibnamefont
  {Westervelt}},\ }\href@noop {} {\bibfield  {journal} {\bibinfo  {journal}
  {Acta Phys. Pol.}\ }\textbf {\bibinfo {volume} {27}},\ \bibinfo {pages} {831}
  (\bibinfo {year} {1965})}\BibitemShut {NoStop}%
\bibitem [{\citenamefont {Hegarty}(1969)}]{Hegarty1969}%
  \BibitemOpen
  \bibfield  {author} {\bibinfo {author} {\bibfnamefont {J.~C.}\ \bibnamefont
  {Hegarty}},\ }\href {https://doi.org/10.1007/BF02711695} {\bibfield
  {journal} {\bibinfo  {journal} {Il Nuovo Cimento B (1965-1970)}\ }\textbf
  {\bibinfo {volume} {61}},\ \bibinfo {pages} {47} (\bibinfo {year}
  {1969})}\BibitemShut {NoStop}%
\bibitem [{\citenamefont {Scully}(1979)}]{PhysRevD.19.3582}%
  \BibitemOpen
  \bibfield  {author} {\bibinfo {author} {\bibfnamefont {M.~O.}\ \bibnamefont
  {Scully}},\ }\href {https://doi.org/10.1103/PhysRevD.19.3582} {\bibfield
  {journal} {\bibinfo  {journal} {Phys. Rev. D}\ }\textbf {\bibinfo {volume}
  {19}},\ \bibinfo {pages} {3582} (\bibinfo {year} {1979})}\BibitemShut
  {NoStop}%
\bibitem [{\citenamefont {R{\"a}tzel}\ \emph {et~al.}(2016)\citenamefont
  {R{\"a}tzel}, \citenamefont {Wilkens},\ and\ \citenamefont
  {Menzel}}]{Rätzel_2016}%
  \BibitemOpen
  \bibfield  {author} {\bibinfo {author} {\bibfnamefont {D.}~\bibnamefont
  {R{\"a}tzel}}, \bibinfo {author} {\bibfnamefont {M.}~\bibnamefont
  {Wilkens}},\ and\ \bibinfo {author} {\bibfnamefont {R.}~\bibnamefont
  {Menzel}},\ }\href {https://doi.org/10.1088/1367-2630/18/2/023009} {\bibfield
   {journal} {\bibinfo  {journal} {New J. Phys.}\ }\textbf {\bibinfo {volume}
  {18}},\ \bibinfo {pages} {023009} (\bibinfo {year} {2016})}\BibitemShut
  {NoStop}%
\bibitem [{\citenamefont {R{\"a}tzel}\ \emph {et~al.}(2017)\citenamefont
  {R{\"a}tzel}, \citenamefont {Wilkens},\ and\ \citenamefont
  {Menzel}}]{PhysRevD.95.084008}%
  \BibitemOpen
  \bibfield  {author} {\bibinfo {author} {\bibfnamefont {D.}~\bibnamefont
  {R{\"a}tzel}}, \bibinfo {author} {\bibfnamefont {M.}~\bibnamefont
  {Wilkens}},\ and\ \bibinfo {author} {\bibfnamefont {R.}~\bibnamefont
  {Menzel}},\ }\href {https://doi.org/10.1103/PhysRevD.95.084008} {\bibfield
  {journal} {\bibinfo  {journal} {Phys. Rev. D}\ }\textbf {\bibinfo {volume}
  {95}},\ \bibinfo {pages} {084008} (\bibinfo {year} {2017})}\BibitemShut
  {NoStop}%
\bibitem [{\citenamefont {Schneiter}\ \emph {et~al.}(2018)\citenamefont
  {Schneiter}, \citenamefont {R{\"a}tzel},\ and\ \citenamefont
  {Braun}}]{Schneiter_2018}%
  \BibitemOpen
  \bibfield  {author} {\bibinfo {author} {\bibfnamefont {F.}~\bibnamefont
  {Schneiter}}, \bibinfo {author} {\bibfnamefont {D.}~\bibnamefont
  {R{\"a}tzel}},\ and\ \bibinfo {author} {\bibfnamefont {D.}~\bibnamefont
  {Braun}},\ }\href {https://doi.org/10.1088/1361-6382/aadc81} {\bibfield
  {journal} {\bibinfo  {journal} {Class. Quantum Gravity}\ }\textbf {\bibinfo
  {volume} {35}},\ \bibinfo {pages} {195007} (\bibinfo {year}
  {2018})}\BibitemShut {NoStop}%
\bibitem [{\citenamefont {Lageyre}\ \emph {et~al.}(2022)\citenamefont
  {Lageyre}, \citenamefont {{d'Humi{\`e}res}},\ and\ \citenamefont
  {Ribeyre}}]{PhysRevD.105.104052}%
  \BibitemOpen
  \bibfield  {author} {\bibinfo {author} {\bibfnamefont {P.}~\bibnamefont
  {Lageyre}}, \bibinfo {author} {\bibfnamefont {E.}~\bibnamefont
  {{d'Humi{\`e}res}}},\ and\ \bibinfo {author} {\bibfnamefont {X.}~\bibnamefont
  {Ribeyre}},\ }\href {https://doi.org/10.1103/PhysRevD.105.104052} {\bibfield
  {journal} {\bibinfo  {journal} {Phys. Rev. D}\ }\textbf {\bibinfo {volume}
  {105}},\ \bibinfo {pages} {104052} (\bibinfo {year} {2022})}\BibitemShut
  {NoStop}%
\bibitem [{\citenamefont {Grishchuk}\ and\ \citenamefont
  {Sazhin}(1973)}]{Grishchuk:1973qz}%
  \BibitemOpen
  \bibfield  {author} {\bibinfo {author} {\bibfnamefont {L.~P.}\ \bibnamefont
  {Grishchuk}}\ and\ \bibinfo {author} {\bibfnamefont {M.~V.}\ \bibnamefont
  {Sazhin}},\ }\href@noop {} {\bibfield  {journal} {\bibinfo  {journal} {Zh.
  Eksp. Teor. Fiz.}\ }\textbf {\bibinfo {volume} {65}},\ \bibinfo {pages} {441}
  (\bibinfo {year} {1973})}\BibitemShut {NoStop}%
\bibitem [{\citenamefont {Grishchuk}\ and\ \citenamefont
  {Sazhin}(1975)}]{1975ZhETF..68.1569G}%
  \BibitemOpen
  \bibfield  {author} {\bibinfo {author} {\bibfnamefont {L.~P.}\ \bibnamefont
  {Grishchuk}}\ and\ \bibinfo {author} {\bibfnamefont {M.~V.}\ \bibnamefont
  {Sazhin}},\ }\href@noop {} {\bibfield  {journal} {\bibinfo  {journal} {Zh.
  Eksp. Teor. Fiz.}\ }\textbf {\bibinfo {volume} {68}},\ \bibinfo {pages}
  {1569} (\bibinfo {year} {1975})}\BibitemShut {NoStop}%
\bibitem [{\citenamefont {Grishchuk}(1977)}]{Leonid_P_Grishchuk_1977}%
  \BibitemOpen
  \bibfield  {author} {\bibinfo {author} {\bibfnamefont {L.~P.}\ \bibnamefont
  {Grishchuk}},\ }\href {https://doi.org/10.1070/PU1977v020n04ABEH005327}
  {\bibfield  {journal} {\bibinfo  {journal} {Sov. Phys. Uspekhi}\ }\textbf
  {\bibinfo {volume} {20}},\ \bibinfo {pages} {319} (\bibinfo {year}
  {1977})}\BibitemShut {NoStop}%
\bibitem [{\citenamefont
  {Grishchuk}(2003)}]{grishchuk2003electromagneticgeneratorsdetectorsgravitational}%
  \BibitemOpen
  \bibfield  {author} {\bibinfo {author} {\bibfnamefont {L.~P.}\ \bibnamefont
  {Grishchuk}},\ }\href@noop {} {\bibinfo {title} {Electromagnetic generators
  and detectors of gravitational waves}} (\bibinfo {year} {2003}),\ \Eprint
  {https://arxiv.org/abs/gr-qc/0306013} {arXiv:gr-qc/0306013} \BibitemShut
  {NoStop}%
\bibitem [{\citenamefont {Spengler}\ \emph {et~al.}(2022)\citenamefont
  {Spengler}, \citenamefont {R{\"a}tzel},\ and\ \citenamefont
  {Braun}}]{Spengler2022}%
  \BibitemOpen
  \bibfield  {author} {\bibinfo {author} {\bibfnamefont {F.}~\bibnamefont
  {Spengler}}, \bibinfo {author} {\bibfnamefont {D.}~\bibnamefont
  {R{\"a}tzel}},\ and\ \bibinfo {author} {\bibfnamefont {D.}~\bibnamefont
  {Braun}},\ }\href {https://doi.org/10.1088/1367-2630/ac5372} {\bibfield
  {journal} {\bibinfo  {journal} {New J. Phys.}\ }\textbf {\bibinfo {volume}
  {24}},\ \bibinfo {pages} {053021} (\bibinfo {year} {2022})}\BibitemShut
  {NoStop}%
\bibitem [{\citenamefont {Gertsenshtein}(1962)}]{Gertsenshtein1962}%
  \BibitemOpen
  \bibfield  {author} {\bibinfo {author} {\bibfnamefont {M.}~\bibnamefont
  {Gertsenshtein}},\ }\href@noop {} {\bibfield  {journal} {\bibinfo  {journal}
  {Sov. Phys. JETP}\ }\textbf {\bibinfo {volume} {14}},\ \bibinfo {pages} {84}
  (\bibinfo {year} {1962})}\BibitemShut {NoStop}%
\bibitem [{\citenamefont {Zel'dovich}\ and\ \citenamefont
  {Novikov}(1983)}]{Zeldovich1983}%
  \BibitemOpen
  \bibfield  {author} {\bibinfo {author} {\bibfnamefont {{\relax Ya}.~B.}\
  \bibnamefont {Zel'dovich}}\ and\ \bibinfo {author} {\bibfnamefont {I.~D.}\
  \bibnamefont {Novikov}},\ }\href@noop {} {\emph {\bibinfo {title}
  {Relativistic astrophysics}}},\ Vol.~\bibinfo {volume} {2}\ (\bibinfo
  {publisher} {The University of Chicago Press},\ \bibinfo {address}
  {Chicago},\ \bibinfo {year} {1983})\BibitemShut {NoStop}%
\bibitem [{\citenamefont {Portilla}\ and\ \citenamefont
  {Lapiedra}(2001)}]{PhysRevD.63.044014}%
  \BibitemOpen
  \bibfield  {author} {\bibinfo {author} {\bibfnamefont {M.}~\bibnamefont
  {Portilla}}\ and\ \bibinfo {author} {\bibfnamefont {R.}~\bibnamefont
  {Lapiedra}},\ }\href {https://doi.org/10.1103/PhysRevD.63.044014} {\bibfield
  {journal} {\bibinfo  {journal} {Phys. Rev. D}\ }\textbf {\bibinfo {volume}
  {63}},\ \bibinfo {pages} {044014} (\bibinfo {year} {2001})}\BibitemShut
  {NoStop}%
\bibitem [{\citenamefont {Kolosnitsyn}\ and\ \citenamefont
  {Rudenko}(2015)}]{Kolosnitsyn_2015}%
  \BibitemOpen
  \bibfield  {author} {\bibinfo {author} {\bibfnamefont {N.~I.}\ \bibnamefont
  {Kolosnitsyn}}\ and\ \bibinfo {author} {\bibfnamefont {V.~N.}\ \bibnamefont
  {Rudenko}},\ }\href {https://doi.org/10.1088/0031-8949/90/7/074059}
  {\bibfield  {journal} {\bibinfo  {journal} {Phys. Scr.}\ }\textbf {\bibinfo
  {volume} {90}},\ \bibinfo {pages} {074059} (\bibinfo {year}
  {2015})}\BibitemShut {NoStop}%
\bibitem [{\citenamefont {Ribeyre}\ and\ \citenamefont
  {Tikhonchuk}()}]{10.1142/9789814374552_0292}%
  \BibitemOpen
  \bibfield  {author} {\bibinfo {author} {\bibfnamefont {X.}~\bibnamefont
  {Ribeyre}}\ and\ \bibinfo {author} {\bibfnamefont {V.}~\bibnamefont
  {Tikhonchuk}},\ }in\ \href {https://doi.org/10.1142/9789814374552_0292}
  {\emph {\bibinfo {booktitle} {The {{Twelfth Marcel Grossmann Meeting}}}}},\
  pp.\ \bibinfo {pages} {1640--1642}\BibitemShut {NoStop}%
\bibitem [{\citenamefont {Gelfer}\ \emph {et~al.}(2016)\citenamefont {Gelfer},
  \citenamefont {Kadlecov{\'a}}, \citenamefont {Klimo}, \citenamefont {Weber},\
  and\ \citenamefont {Korn}}]{10.1063/1.4962520}%
  \BibitemOpen
  \bibfield  {author} {\bibinfo {author} {\bibfnamefont {E.~G.}\ \bibnamefont
  {Gelfer}}, \bibinfo {author} {\bibfnamefont {H.}~\bibnamefont
  {Kadlecov{\'a}}}, \bibinfo {author} {\bibfnamefont {O.}~\bibnamefont
  {Klimo}}, \bibinfo {author} {\bibfnamefont {S.}~\bibnamefont {Weber}},\ and\
  \bibinfo {author} {\bibfnamefont {G.}~\bibnamefont {Korn}},\ }\href
  {https://doi.org/10.1063/1.4962520} {\bibfield  {journal} {\bibinfo
  {journal} {Phys. Plasmas}\ }\textbf {\bibinfo {volume} {23}},\ \bibinfo
  {pages} {093107} (\bibinfo {year} {2016})}\BibitemShut {NoStop}%
\bibitem [{\citenamefont {Kadlecov{\'a}}\ \emph {et~al.}(2017)\citenamefont
  {Kadlecov{\'a}}, \citenamefont {Klimo}, \citenamefont {Weber},\ and\
  \citenamefont {Korn}}]{Kadlecova2017}%
  \BibitemOpen
  \bibfield  {author} {\bibinfo {author} {\bibfnamefont {H.}~\bibnamefont
  {Kadlecov{\'a}}}, \bibinfo {author} {\bibfnamefont {O.}~\bibnamefont
  {Klimo}}, \bibinfo {author} {\bibfnamefont {S.}~\bibnamefont {Weber}},\ and\
  \bibinfo {author} {\bibfnamefont {G.}~\bibnamefont {Korn}},\ }\href
  {https://doi.org/10.1140/epjd/e2017-70586-y} {\bibfield  {journal} {\bibinfo
  {journal} {Eur. Phys. J. D}\ }\textbf {\bibinfo {volume} {71}},\ \bibinfo
  {pages} {89} (\bibinfo {year} {2017})}\BibitemShut {NoStop}%
\bibitem [{\citenamefont {Bonnor}(1969)}]{Bonnor1969}%
  \BibitemOpen
  \bibfield  {author} {\bibinfo {author} {\bibfnamefont {W.~B.}\ \bibnamefont
  {Bonnor}},\ }\href {https://doi.org/10.1007/BF01645484} {\bibfield  {journal}
  {\bibinfo  {journal} {Commun. Math. Phys.}\ }\textbf {\bibinfo {volume}
  {13}},\ \bibinfo {pages} {163} (\bibinfo {year} {1969})}\BibitemShut
  {NoStop}%
\bibitem [{\citenamefont {Griffiths}(2016)}]{griffiths2016colliding}%
  \BibitemOpen
  \bibfield  {author} {\bibinfo {author} {\bibfnamefont {J.}~\bibnamefont
  {Griffiths}},\ }in\ \href@noop {} {\emph {\bibinfo {booktitle} {Colliding
  plane waves in general relativity}}}\ (\bibinfo  {publisher} {Dover
  Publications},\ \bibinfo {year} {2016})\ Chap.~\bibinfo {chapter}
  {4}\BibitemShut {NoStop}%
\bibitem [{\citenamefont {Brinkmann}(1923)}]{doi:10.1073/pnas.9.1.1}%
  \BibitemOpen
  \bibfield  {author} {\bibinfo {author} {\bibfnamefont {H.~W.}\ \bibnamefont
  {Brinkmann}},\ }\href {https://doi.org/10.1073/pnas.9.1.1} {\bibfield
  {journal} {\bibinfo  {journal} {Proc. Natl. Acad. Sci. U.S.A.}\ }\textbf
  {\bibinfo {volume} {9}},\ \bibinfo {pages} {1} (\bibinfo {year}
  {1923})}\BibitemShut {NoStop}%
\bibitem [{\citenamefont {Peres}(1959)}]{PhysRevLett.3.571}%
  \BibitemOpen
  \bibfield  {author} {\bibinfo {author} {\bibfnamefont {A.}~\bibnamefont
  {Peres}},\ }\href {https://doi.org/10.1103/PhysRevLett.3.571} {\bibfield
  {journal} {\bibinfo  {journal} {Phys. Rev. Lett.}\ }\textbf {\bibinfo
  {volume} {3}},\ \bibinfo {pages} {571} (\bibinfo {year} {1959})}\BibitemShut
  {NoStop}%
\bibitem [{\citenamefont {{van
  Holten}}(2011)}]{https://doi.org/10.1002/prop.201000088}%
  \BibitemOpen
  \bibfield  {author} {\bibinfo {author} {\bibfnamefont {J.}~\bibnamefont {{van
  Holten}}},\ }\href {https://doi.org/10.1002/prop.201000088} {\bibfield
  {journal} {\bibinfo  {journal} {Fortschr. Phys.}\ }\textbf {\bibinfo {volume}
  {59}},\ \bibinfo {pages} {284} (\bibinfo {year} {2011})}\BibitemShut
  {NoStop}%
\bibitem [{\citenamefont {Rosen}(1937)}]{rosen1937plane}%
  \BibitemOpen
  \bibfield  {author} {\bibinfo {author} {\bibfnamefont {N.}~\bibnamefont
  {Rosen}},\ }\href@noop {} {\bibfield  {journal} {\bibinfo  {journal} {Phys.
  Z. Sowjetunion}\ }\textbf {\bibinfo {volume} {12}},\ \bibinfo {pages} {366}
  (\bibinfo {year} {1937})}\BibitemShut {NoStop}%
\bibitem [{\citenamefont {Bell}\ and\ \citenamefont
  {Szekeres}(1974)}]{Bell1974}%
  \BibitemOpen
  \bibfield  {author} {\bibinfo {author} {\bibfnamefont {P.}~\bibnamefont
  {Bell}}\ and\ \bibinfo {author} {\bibfnamefont {P.}~\bibnamefont
  {Szekeres}},\ }\href {https://doi.org/10.1007/BF00770217} {\bibfield
  {journal} {\bibinfo  {journal} {Gen. Relativ. Gravit.}\ }\textbf {\bibinfo
  {volume} {5}},\ \bibinfo {pages} {275} (\bibinfo {year} {1974})}\BibitemShut
  {NoStop}%
\bibitem [{\citenamefont {Bini}\ \emph {et~al.}(2014)\citenamefont {Bini},
  \citenamefont {Geralico}, \citenamefont {Haney},\ and\ \citenamefont
  {Ortolan}}]{PhysRevD.89.104049}%
  \BibitemOpen
  \bibfield  {author} {\bibinfo {author} {\bibfnamefont {D.}~\bibnamefont
  {Bini}}, \bibinfo {author} {\bibfnamefont {A.}~\bibnamefont {Geralico}},
  \bibinfo {author} {\bibfnamefont {M.}~\bibnamefont {Haney}},\ and\ \bibinfo
  {author} {\bibfnamefont {A.}~\bibnamefont {Ortolan}},\ }\href
  {https://doi.org/10.1103/PhysRevD.89.104049} {\bibfield  {journal} {\bibinfo
  {journal} {Phys. Rev. D}\ }\textbf {\bibinfo {volume} {89}},\ \bibinfo
  {pages} {104049} (\bibinfo {year} {2014})}\BibitemShut {NoStop}%
\bibitem [{\citenamefont {Aichelburg}\ and\ \citenamefont
  {Sexl}(1971)}]{Aichelburg1971}%
  \BibitemOpen
  \bibfield  {author} {\bibinfo {author} {\bibfnamefont {P.~C.}\ \bibnamefont
  {Aichelburg}}\ and\ \bibinfo {author} {\bibfnamefont {R.~U.}\ \bibnamefont
  {Sexl}},\ }\href {https://doi.org/10.1007/BF00758149} {\bibfield  {journal}
  {\bibinfo  {journal} {Gen. Relativ. Gravit.}\ }\textbf {\bibinfo {volume}
  {2}},\ \bibinfo {pages} {303} (\bibinfo {year} {1971})}\BibitemShut {NoStop}%
\bibitem [{\citenamefont {Bonnor}(2009)}]{Bonnor2009}%
  \BibitemOpen
  \bibfield  {author} {\bibinfo {author} {\bibfnamefont {W.~B.}\ \bibnamefont
  {Bonnor}},\ }\href {https://doi.org/10.1007/s10714-008-0655-z} {\bibfield
  {journal} {\bibinfo  {journal} {Gen. Relativ. Gravit.}\ }\textbf {\bibinfo
  {volume} {41}},\ \bibinfo {pages} {77} (\bibinfo {year} {2009})}\BibitemShut
  {NoStop}%
\bibitem [{Note1()}]{Note1}%
  \BibitemOpen
  \bibinfo {note} {Discontinuities in the second-order derivatives are
  expected, as the Einstein equation is a second-order partial differential
  equation, and the stress-energy tensor is discontinuous at the border of the
  pulse.}\BibitemShut {Stop}%
\bibitem [{\citenamefont {Lichnerowicz}(1955)}]{lichnerowicz1955théories}%
  \BibitemOpen
  \bibfield  {author} {\bibinfo {author} {\bibfnamefont {A.}~\bibnamefont
  {Lichnerowicz}},\ }in\ \href@noop {} {\emph {\bibinfo {booktitle}
  {Th{\'e}ories relativistes de la gravitation et de
  l'{\'e}lectromagn{\'e}tisme: relativit{\'e} g{\'e}n{\'e}rale et th{\'e}ories
  unitaires}}},\ \bibinfo {series and number} {Collection d'ouvrages de
  math{\'e}matiques {\`a} l'usage des physiciens}\ (\bibinfo  {publisher}
  {Masson},\ \bibinfo {year} {1955})\ Chap.\ \bibinfo {chapter} {I,
  III}\BibitemShut {NoStop}%
\bibitem [{Sup()}]{Supplemental}%
  \BibitemOpen
  \href@noop {} {}\bibinfo {note} {See Supplemental Material for a detailed
  derivation of the spacetime associated with light pulses and the trajectories
  of particles perturbed by the gravitational shock waves.}\BibitemShut {Stop}%
\bibitem [{\citenamefont {Gori}(1994)}]{GORI1994335}%
  \BibitemOpen
  \bibfield  {author} {\bibinfo {author} {\bibfnamefont {F.}~\bibnamefont
  {Gori}},\ }\href {https://doi.org/10.1016/0030-4018(94)90342-5} {\bibfield
  {journal} {\bibinfo  {journal} {Opt. Commun.}\ }\textbf {\bibinfo {volume}
  {107}},\ \bibinfo {pages} {335} (\bibinfo {year} {1994})}\BibitemShut
  {NoStop}%
\bibitem [{Note2()}]{Note2}%
  \BibitemOpen
  \bibinfo {note} {The domain of inverse error function is $(-1,1)$. In this
  paper, we assume $\epsilon \ll \protect \bar {r}_\protect \text
  {p}^2/(\protect \bar {u}_\protect \text {b}-\protect \bar {u}_\protect \text
  {a})^2$, which ensures that the transformation (\ref {vr_pg_vr_xieta}) is
  well-defined throughout the entire region $\protect \mathcal {R}_\protect
  \text {g}$. If this condition is not satisfied, it becomes necessary to
  define subregions $\protect \mathcal {D}_\xi \subset \protect \mathcal
  {R}_\protect \text {g}$ as $| u - \protect \bar {u}_\xi | / r_{\xi \protect
  \text {g}} < \protect \sqrt {\pi /2 \epsilon }$ and assume that the
  coordinate systems $(u, v_{\xi \protect \text {g}}, r_{\xi \protect \text
  {g}}, \varphi )$ cover the subregions $\protect \mathcal {R}_\xi \cup
  \protect \mathcal {D}_\xi \subset \protect \mathcal {R}_\xi \cup \protect
  \mathcal {R}_\protect \text {g}$ instead of $\protect \mathcal {R}_\xi \cup
  \protect \mathcal {R}_\protect \text {g}$. This modification does not affect
  the results of this paper, as the coordinate systems $(u, v_{\protect \text
  {p} \protect \text {g}}, r_{\protect \text {p} \protect \text {g}}, \varphi
  )$ and $(u, v_{\xi \eta }, r_{\xi \eta }, \varphi )$ still provide a complete
  atlas covering both the interior and exterior regions.}\BibitemShut {Stop}%
\bibitem [{\citenamefont {Pikovski}\ \emph {et~al.}(2012)\citenamefont
  {Pikovski}, \citenamefont {Vanner}, \citenamefont {Aspelmeyer}, \citenamefont
  {Kim},\ and\ \citenamefont {Brukner}}]{Pikovski2012}%
  \BibitemOpen
  \bibfield  {author} {\bibinfo {author} {\bibfnamefont {I.}~\bibnamefont
  {Pikovski}}, \bibinfo {author} {\bibfnamefont {M.~R.}\ \bibnamefont
  {Vanner}}, \bibinfo {author} {\bibfnamefont {M.}~\bibnamefont {Aspelmeyer}},
  \bibinfo {author} {\bibfnamefont {M.~S.}\ \bibnamefont {Kim}},\ and\ \bibinfo
  {author} {\bibfnamefont {{\v C}.}~\bibnamefont {Brukner}},\ }\href
  {https://doi.org/10.1038/nphys2262} {\bibfield  {journal} {\bibinfo
  {journal} {Nat. Phys.}\ }\textbf {\bibinfo {volume} {8}},\ \bibinfo {pages}
  {393} (\bibinfo {year} {2012})}\BibitemShut {NoStop}%
\bibitem [{\citenamefont {Schm{\"o}le}\ \emph {et~al.}(2016)\citenamefont
  {Schm{\"o}le}, \citenamefont {Dragosits}, \citenamefont {Hepach},\ and\
  \citenamefont {Aspelmeyer}}]{Schmöle_2016}%
  \BibitemOpen
  \bibfield  {author} {\bibinfo {author} {\bibfnamefont {J.}~\bibnamefont
  {Schm{\"o}le}}, \bibinfo {author} {\bibfnamefont {M.}~\bibnamefont
  {Dragosits}}, \bibinfo {author} {\bibfnamefont {H.}~\bibnamefont {Hepach}},\
  and\ \bibinfo {author} {\bibfnamefont {M.}~\bibnamefont {Aspelmeyer}},\
  }\href {https://doi.org/10.1088/0264-9381/33/12/125031} {\bibfield  {journal}
  {\bibinfo  {journal} {Class. Quantum Gravity}\ }\textbf {\bibinfo {volume}
  {33}},\ \bibinfo {pages} {125031} (\bibinfo {year} {2016})}\BibitemShut
  {NoStop}%
\bibitem [{\citenamefont {Belenchia}\ \emph {et~al.}(2018)\citenamefont
  {Belenchia}, \citenamefont {Wald}, \citenamefont {Giacomini}, \citenamefont
  {{Castro-Ruiz}}, \citenamefont {Brukner},\ and\ \citenamefont
  {Aspelmeyer}}]{PhysRevD.98.126009}%
  \BibitemOpen
  \bibfield  {author} {\bibinfo {author} {\bibfnamefont {A.}~\bibnamefont
  {Belenchia}}, \bibinfo {author} {\bibfnamefont {R.~M.}\ \bibnamefont {Wald}},
  \bibinfo {author} {\bibfnamefont {F.}~\bibnamefont {Giacomini}}, \bibinfo
  {author} {\bibfnamefont {E.}~\bibnamefont {{Castro-Ruiz}}}, \bibinfo {author}
  {\bibfnamefont {{\v C}.}~\bibnamefont {Brukner}},\ and\ \bibinfo {author}
  {\bibfnamefont {M.}~\bibnamefont {Aspelmeyer}},\ }\href
  {https://doi.org/10.1103/PhysRevD.98.126009} {\bibfield  {journal} {\bibinfo
  {journal} {Phys. Rev. D}\ }\textbf {\bibinfo {volume} {98}},\ \bibinfo
  {pages} {126009} (\bibinfo {year} {2018})}\BibitemShut {NoStop}%
\bibitem [{\citenamefont {Marletto}\ and\ \citenamefont
  {Vedral}(2017)}]{PhysRevLett.119.240402}%
  \BibitemOpen
  \bibfield  {author} {\bibinfo {author} {\bibfnamefont {C.}~\bibnamefont
  {Marletto}}\ and\ \bibinfo {author} {\bibfnamefont {V.}~\bibnamefont
  {Vedral}},\ }\href {https://doi.org/10.1103/PhysRevLett.119.240402}
  {\bibfield  {journal} {\bibinfo  {journal} {Phys. Rev. Lett.}\ }\textbf
  {\bibinfo {volume} {119}},\ \bibinfo {pages} {240402} (\bibinfo {year}
  {2017})}\BibitemShut {NoStop}%
\bibitem [{\citenamefont {Lewenstein}\ \emph {et~al.}(2021)\citenamefont
  {Lewenstein}, \citenamefont {Ciappina}, \citenamefont {Pisanty},
  \citenamefont {{Rivera-Dean}}, \citenamefont {Stammer}, \citenamefont
  {Lamprou},\ and\ \citenamefont {Tzallas}}]{Lewenstein2021}%
  \BibitemOpen
  \bibfield  {author} {\bibinfo {author} {\bibfnamefont {M.}~\bibnamefont
  {Lewenstein}}, \bibinfo {author} {\bibfnamefont {M.~F.}\ \bibnamefont
  {Ciappina}}, \bibinfo {author} {\bibfnamefont {E.}~\bibnamefont {Pisanty}},
  \bibinfo {author} {\bibfnamefont {J.}~\bibnamefont {{Rivera-Dean}}}, \bibinfo
  {author} {\bibfnamefont {P.}~\bibnamefont {Stammer}}, \bibinfo {author}
  {\bibfnamefont {{\relax Th}.}~\bibnamefont {Lamprou}},\ and\ \bibinfo
  {author} {\bibfnamefont {P.}~\bibnamefont {Tzallas}},\ }\href
  {https://doi.org/10.1038/s41567-021-01317-w} {\bibfield  {journal} {\bibinfo
  {journal} {Nat. Phys.}\ }\textbf {\bibinfo {volume} {17}},\ \bibinfo {pages}
  {1104} (\bibinfo {year} {2021})}\BibitemShut {NoStop}%
\bibitem [{\citenamefont {Chen}\ and\ \citenamefont
  {Giacomini}(2024)}]{chen2024quantumeffectsgravitynewton}%
  \BibitemOpen
  \bibfield  {author} {\bibinfo {author} {\bibfnamefont {L.-Q.}\ \bibnamefont
  {Chen}}\ and\ \bibinfo {author} {\bibfnamefont {F.}~\bibnamefont
  {Giacomini}},\ }\href@noop {} {\bibinfo {title} {Quantum effects in gravity
  beyond the {{Newton}} potential from a delocalised quantum source}} (\bibinfo
  {year} {2024}),\ \Eprint {https://arxiv.org/abs/2402.10288} {arXiv:2402.10288
  [quant-ph]} \BibitemShut {NoStop}%
\end{thebibliography}%


\begin{thebibliography}{3}%
\makeatletter
\providecommand \@ifxundefined [1]{%
 \@ifx{#1\undefined}
}%
\providecommand \@ifnum [1]{%
 \ifnum #1\expandafter \@firstoftwo
 \else \expandafter \@secondoftwo
 \fi
}%
\providecommand \@ifx [1]{%
 \ifx #1\expandafter \@firstoftwo
 \else \expandafter \@secondoftwo
 \fi
}%
\providecommand \natexlab [1]{#1}%
\providecommand \enquote  [1]{``#1''}%
\providecommand \bibnamefont  [1]{#1}%
\providecommand \bibfnamefont [1]{#1}%
\providecommand \citenamefont [1]{#1}%
\providecommand \href@noop [0]{\@secondoftwo}%
\providecommand \href [0]{\begingroup \@sanitize@url \@href}%
\providecommand \@href[1]{\@@startlink{#1}\@@href}%
\providecommand \@@href[1]{\endgroup#1\@@endlink}%
\providecommand \@sanitize@url [0]{\catcode `\\12\catcode `\$12\catcode
  `\&12\catcode `\#12\catcode `\^12\catcode `\_12\catcode `\%12\relax}%
\providecommand \@@startlink[1]{}%
\providecommand \@@endlink[0]{}%
\providecommand \url  [0]{\begingroup\@sanitize@url \@url }%
\providecommand \@url [1]{\endgroup\@href {#1}{\urlprefix }}%
\providecommand \urlprefix  [0]{URL }%
\providecommand \Eprint [0]{\href }%
\providecommand \doibase [0]{https://doi.org/}%
\providecommand \selectlanguage [0]{\@gobble}%
\providecommand \bibinfo  [0]{\@secondoftwo}%
\providecommand \bibfield  [0]{\@secondoftwo}%
\providecommand \translation [1]{[#1]}%
\providecommand \BibitemOpen [0]{}%
\providecommand \bibitemStop [0]{}%
\providecommand \bibitemNoStop [0]{.\EOS\space}%
\providecommand \EOS [0]{\spacefactor3000\relax}%
\providecommand \BibitemShut  [1]{\csname bibitem#1\endcsname}%
\let\auto@bib@innerbib\@empty
\bibitem [{\citenamefont {{van
  Holten}}(2011)}]{https://doi.org/10.1002/prop.201000088}%
  \BibitemOpen
  \bibfield  {author} {\bibinfo {author} {\bibfnamefont {J.}~\bibnamefont {{van
  Holten}}},\ }\href {https://doi.org/10.1002/prop.201000088} {\bibfield
  {journal} {\bibinfo  {journal} {Fortschr. Phys.}\ }\textbf {\bibinfo {volume}
  {59}},\ \bibinfo {pages} {284} (\bibinfo {year} {2011})}\BibitemShut
  {NoStop}%
\bibitem [{\citenamefont {Bini}\ \emph {et~al.}(2014)\citenamefont {Bini},
  \citenamefont {Geralico}, \citenamefont {Haney},\ and\ \citenamefont
  {Ortolan}}]{PhysRevD.89.104049}%
  \BibitemOpen
  \bibfield  {author} {\bibinfo {author} {\bibfnamefont {D.}~\bibnamefont
  {Bini}}, \bibinfo {author} {\bibfnamefont {A.}~\bibnamefont {Geralico}},
  \bibinfo {author} {\bibfnamefont {M.}~\bibnamefont {Haney}},\ and\ \bibinfo
  {author} {\bibfnamefont {A.}~\bibnamefont {Ortolan}},\ }\href
  {https://doi.org/10.1103/PhysRevD.89.104049} {\bibfield  {journal} {\bibinfo
  {journal} {Phys. Rev. D}\ }\textbf {\bibinfo {volume} {89}},\ \bibinfo
  {pages} {104049} (\bibinfo {year} {2014})}\BibitemShut {NoStop}%
\bibitem [{Note1()}]{Note1}%
  \BibitemOpen
  \bibinfo {note} {Here, $\protect \bar {u}_\xi $ can assume any value in
  $\protect \mathbb {R}$. Conversely, in the main paper, it is constrained to
  be either $\protect \bar {u}_\protect \text {a}$ or $\protect \bar
  {u}_\protect \text {b}$.}\BibitemShut {Stop}%
\end{thebibliography}%

\end{document}


\title{Supplemental Material: Cumulative effects of laser-generated gravitational shock waves}

\author{Riccardo Falcone}
\affiliation{Department of Physics, University of Sapienza, Piazzale Aldo Moro 5, 00185 Rome, Italy}

\author{Claudio Conti}
\affiliation{Department of Physics, University of Sapienza, Piazzale Aldo Moro 5, 00185 Rome, Italy}

\maketitle

\section{Spacetime of a single pulse}\label{Spacetime_of_a_single_pulse}

In this section, we consider a cylindrical energy density propagating at the speed of light. The spacetime is divided into four distinct regions, corresponding to the interior of the cylinder ($\mathcal{R}_\text{p}$), the surrounding Gravitational Wave (GW) ($\mathcal{R}_\text{g}$) and the exterior regions before ($\mathcal{R}_\text{a}$) and after ($\mathcal{R}_\text{b}$) the passage of the pulse and the GW. Cylindrical symmetry is assumed for both $\mathcal{R}_\text{p}$ and $\mathcal{R}_\text{g}$. In the remaining regions ($\mathcal{R}_\text{a}$ and $\mathcal{R}_\text{b}$), the metric is taken to be Minkowski due to the absence of any gravitational source. We derive the global spacetime structure by solving the Einstein equation, with continuity conditions imposed at each boundary.

\subsection{Light pulse}\label{Light_pulse}

In this subsection, we examine the interior region of the pulse $\mathcal{R}_\text{p}$ by using two distinct coordinate systems: Brinkmann coordinates and Rosen coordinates. The Brinkmann coordinates are adapted to the cylindrical symmetry of $\mathcal{R}_\text{p}$, whereas the Rosen coordinates are locally Minkowskian near hyperplanes of constant light-cone coordinate $u$.

\subsubsection{Brinkmann-pulse coordinates}\label{Brinkmann_pulse_coordinates}

Assuming a constant energy density traveling with the speed of light, the spacetime can be locally represented through the Brinkmann coordinates $(t_{\text{p}\text{g}}, z_{\text{p}\text{g}}, r_{\text{p}\text{g}}, \varphi)$ \cite{https://doi.org/10.1002/prop.201000088}, with the metric $g_{\mu\nu}$ given by
\begin{equation}\label{g_p_tzr_B}
\left. ds^2 \right|_{\mathcal{R}_\text{p}} = - \frac{1}{2} \kappa  \mathcal{E} r_{\text{p}\text{g}}^2 \left(c dt_{\text{p}\text{g}} - dz_{\text{p}\text{g}} \right)^2 - c^2 dt_{\text{p}\text{g}}^2  + dz_{\text{p}\text{g}}^2 + dr_{\text{p}\text{g}}^2 +r_{\text{p}\text{g}}^2 d\varphi^2.
\end{equation}
Here, $\kappa$ is the Einstein gravitational constant, $c$ is the speed of light and $\mathcal{E}$ is the energy density. In this coordinate system, $t_{\text{p}\text{g}}$ appears as the time coordinate corresponding to the stationary symmetry generated by the Killing vector field $\partial_{t_{\text{p}\text{g}}}$, whereas  $z_{\text{p}\text{g}}$ and $\varphi $ act as cylindrical coordinates associated with the cylindrical symmetry generated by the Killing vector fields $\partial_{z_{\text{p}\text{g}}}$ and $\partial_{\varphi}$.

The metric (\ref{g_p_tzr_B}) satisfies the Einstein equation $R_{\mu\nu} - R g_{\mu\nu}/2 = \kappa T_{\mu\nu}$, where $R_{\mu\nu}$ is the Ricci curvature tensor, $R$ is the scalar curvature and $T_{\mu\nu}$ is the stress-energy tensor with the following nonzero components: $T_{00} = c^2 \mathcal{E}$, $T_{01} = T_{10} = -c \mathcal{E}$, and $T_{11} = \mathcal{E}$. As a type $(2,0)$ tensor, $T$ can be expressed as
\begin{equation}\label{T_p_tzr_B}
\left. T \right|_{\mathcal{R}_\text{p}} = \frac{\mathcal{E} }{c^2} \partial_{t_{\text{p}\text{g}}} \otimes \partial_{t_{\text{p}\text{g}}} + \frac{\mathcal{E}}{c} \partial_{t_{\text{p}\text{g}}} \otimes \partial_{z_{\text{p}\text{g}}} + \frac{\mathcal{E} }{c} \partial_{z_{\text{p}\text{g}}} \otimes \partial_{t_{\text{p}\text{g}}} + \mathcal{E} \partial_{z_{\text{p}\text{g}}} \otimes \partial_{z_{\text{p}\text{g}}} .
\end{equation}
This stress-energy tensor represents a uniform energy density $\mathcal{E}$ traveling at the speed of light along the $z_{\text{p}\text{g}}$ direction.

A practical realization of Eq.\ (\ref{T_p_tzr_B}) can be achieved by using a circularly polarized monochromatic plane wave described by the electromagnetic four-potential
\begin{equation}\label{A_p_tzr_B}
A = \mathcal{A} \left[ \cos(k c t_{\text{p}\text{g}} - k z_{\text{p}\text{g}}-\varphi) dr_{\text{p}\text{g}} + \sin(k c t_{\text{p}\text{g}} - k z_{\text{p}\text{g}}-\varphi) r_{\text{p}\text{g}} d\varphi \right],
\end{equation}
where $\mathcal{A}$ is the amplitude and $k$ is the wave number. By computing the corresponding electromagnetic stress-energy tensor
\begin{equation}
T^{\mu\nu} = - \frac{1}{\mu_0} \left( F^{\mu\rho} F_\rho{}^\nu + \frac{1}{4} g^{\mu\nu} F_{\rho\sigma} F^{\rho\sigma}\right),
\end{equation}
with $F_{\mu\nu} = \partial_\mu A_\nu - \partial_\nu A_\mu$ as the electromagnetic field tensor and $\mu_0$ as the vacuum permeability, we recover Eq.~(\ref{T_p_tzr_B}) with $\mathcal{E} = \mathcal{A}^2 k^2 /\mu_0$. It is also possible to check that $F_{\mu\nu}$ and $g_{\mu\nu}$ satisfy the Maxwell equation in curved spacetime $\nabla_\mu F^{\mu\nu} = 0$. From Eq.\ (\ref{T_p_tzr_B}), the Poynting vector is found to be $S = \mathcal{E} \partial_{z_{\text{p}\text{g}}}$. Accordingly, the intensity of the electromagnetic field is $I = c \mathcal{E}$.

Expressed in light-cone coordinates $u=(ct_{\text{p}\text{g}}-z_{\text{p}\text{g}})/\sqrt{2}$ and $v_{\text{p}\text{g}}=(ct_{\text{p}\text{g}}+z_{\text{p}\text{g}})/\sqrt{2}$, Eq.\ (\ref{g_p_tzr_B}) becomes
\begin{equation}\label{g_p_vr_pg}
\left. ds^2 \right|_{\mathcal{R}_\text{p}} = - \kappa  \mathcal{E} r_{\text{p}\text{g}}^2 du^2 -2du dv_{\text{p}\text{g}} + dr_{\text{p}\text{g}}^2 +r_{\text{p}\text{g}}^2 d\varphi^2.
\end{equation}
Analogously, the stress-energy tensor (\ref{T_p_tzr_B}) becomes $\left. T \right|_{\mathcal{R}_\text{p}} = \mathcal{E} \partial_{v_{\text{p}\text{g}}} \otimes \partial_{v_{\text{p}\text{g}}}$. In this coordinate system, we assume that the region $\mathcal{R}_\text{p}$ confining the pulse is a cylinder defined by $u \in [\bar{u}_\text{a}, \bar{u}_\text{b}]$ and $r_{\text{p}\text{g}} \in [0, \bar{r}_\text{p}]$ for some $\bar{u}_\text{a}$, $\bar{u}_\text{b}$ and $\bar{r}_\text{p}$.

\subsubsection{Rosen-pulse coordinates}\label{Rosen_pulse_coordinates}

As an alternative to the Brinkmann coordinates $(u, v_{\text{p}\text{g}}, r_{\text{p}\text{g}}, \varphi)$, the Rosen coordinate system $(u,v_{\xi\text{p}},r_{\xi\text{p}},\varphi)$ \cite{PhysRevD.89.104049} can be adopted to describe the region $\mathcal{R}_\text{p}$, with the metric $g_{\mu\nu}$ expressed as
\begin{equation}\label{g_p_vr_xip}
\left.  ds^2 \right|_{\mathcal{R}_\text{p} \cap \mathcal{D}_\xi} = - 2 du dv_{\xi\text{p}}+ \left\lbrace \cos \left[\sqrt{\kappa \mathcal{E}} (u- \bar{u}_\xi) \right] \right\rbrace^2 \left( dr_{\xi\text{p}}^2 + r_{\xi\text{p}}^2 d\varphi^2 \right).
\end{equation}
Here, the parameter $\bar{u}_\xi$ can take any real value \footnote{Here, $\bar{u}_\xi$ can assume any value in $\mathbb{R}$. Conversely, in the main paper, it is constrained to be either $\bar{u}_\text{a}$ or $\bar{u}_\text{b}$.}, resulting in a family of coordinate systems $(u,v_{\xi\text{p}},r_{\xi\text{p}},\varphi)$, each corresponding to a different phase $\bar{u}_\xi$ in Eq.\ (\ref{g_p_vr_xip}). For any given $\bar{u}_\xi$, the region $\mathcal{D}_\xi$ specifies where the metric remains non-singular. By convention, we choose the neighborhood of $u = \bar{u}_\xi$, corresponding to $\sqrt{\kappa \mathcal{E}} |u-\bar{u}_\xi|<\pi/2$.

The transformation between the two coordinate systems $(u,v_{\text{p}\text{g}},r_{\text{p}\text{g}},\varphi)$ and $(u,v_{\xi\text{p}},r_{\xi\text{p}},\varphi)$ is given by
\begin{equation}\label{vr_pg_vr_xip}
\left( u, v_{\text{p}\text{g}}, r_{\text{p}\text{g}}, \varphi \right) = \left( u, v_{\xi \text{p}} + V_\text{p} \left(u-\bar{u}_\xi ,r_{\xi \text{p}} \right), R_\text{p} \left(u-\bar{u}_\xi ,r_{\xi \text{p}} \right), \varphi \right),
\end{equation}
where $V_{\text{p}} (u,r)$ and $R_{\text{p}} (u,r)$ are defined as
\begin{align}\label{VR_p}
& V_{\text{p}} (u,r) = - \frac{1}{4}  \sqrt{\kappa \mathcal{E}} \sin \left( 2 \sqrt{\kappa \mathcal{E}} u \right) r^2,  & R_\text{p} (u ,r ) = \cos \left(\sqrt{\kappa \mathcal{E}} u \right) r.
\end{align}
This can be verified by substituting Eq.\ (\ref{vr_pg_vr_xip}) into the metric (\ref{g_p_vr_pg}), resulting in
\begin{align}\label{g_p_vr_xip_RV}
& \left. ds^2 \right|_{\mathcal{R}_\text{p}} = \left\lbrace \left[ \dot{R}_{\text{p}}\left(u-\bar{u}_{\xi },r_{\xi  \text{p}}\right) \right]^2 -\kappa \mathcal{E} R_{\text{p}}^2\left(u-\bar{u}_{\xi },r_{\xi  \text{p}}\right) - 2 \dot{V}_{\text{p}}\left(u-\bar{u}_{\xi },r_{\xi  \text{p}}\right) \right\rbrace d^2 u -2 du dv_{\xi  \text{p}} \nonumber \\
& + 2 \left[ \dot{R}_{\text{p}}\left(u-\bar{u}_{\xi },r_{\xi  \text{p}}\right) R_{\text{p}}'\left(u-\bar{u}_{\xi },r_{\xi  \text{p}}\right)-V_{\text{p}}'\left(u-\bar{u}_{\xi },r_{\xi  \text{p}}\right) \right] du dr_{\xi  \text{p}} + \left[ R_{\text{p}}^{\prime }\left(u-\bar{u}_{\xi },r_{\xi  \text{p}}\right)\right]^2 dr_{\xi  \text{p}}^2 + R_{\text{p}}^2\left(u-\bar{u}_{\xi },r_{\xi  \text{p}}\right) d\varphi^2,
\end{align}
with $\dot{R}_{\text{p}} (u,r ) = \partial_u R_{\text{p}} (u,r )$, $\dot{V}_{\text{p}} (u,r ) = \partial_u V_{\text{p}} (u,r )$, $R'_{\text{p}} (u,r ) = \partial_r R_{\text{p}} (u,r )$, $V'_{\text{p}} (u,r ) = \partial_r V_{\text{p}} (u,r )$. By using Eq.\ (\ref{VR_p}), one can compute these derivatives and confirm that the result matches Eq.\ (\ref{g_p_vr_xip}).

Notice that around $u = \bar{u}_\xi$, the metric can be approximated as
\begin{equation}\label{g_p_vr_xip_u_xi}
\left.  ds^2 \right|_{\mathcal{R}_\text{p} \cap \mathcal{D}_\xi} = - 2 du dv_{\xi\text{p}} + \left[ 1+ O \left( (u- \bar{u}_{\xi\text{p}})^2 \right) \right] \left( dr_{\xi\text{p}}^2 + r_{\xi\text{p}}^2 d\varphi^2 \right).
\end{equation}
This implies that $(u,v_{\xi\text{p}},r_{\xi\text{p}},\varphi)$ is a Fermi coordinate system at any point in $u = \bar{u}_\xi$ within $\mathcal{R}_\text{p}$.

By selecting $\bar{u}_\xi=\bar{u}_\text{a}$ and $\bar{u}_\xi=\bar{u}_\text{b}$, we define the particular Rosen coordinate systems $(u,v_{\text{ap}},r_{\text{ap}},\varphi)$ and $(u,v_{\text{bp}},r_{\text{bp}},\varphi)$, such that
\begin{align}\label{g_p_vr_abp}
& \left.  ds^2 \right|_{\mathcal{R}_\text{p} \cap \mathcal{D}_\text{a}} = - 2 du dv_\text{ap}+ \left\lbrace \cos \left[\sqrt{\kappa \mathcal{E}} (u- \bar{u}_\text{a}) \right] \right\rbrace^2 \left( dr_\text{ap}^2 + r_\text{ap}^2 d\varphi^2 \right), \nonumber \\
 & \left.  ds^2 \right|_{\mathcal{R}_\text{p} \cap \mathcal{D}_\text{b}} = - 2 du dv_\text{bp}+ \left\lbrace \cos \left[\sqrt{\kappa \mathcal{E}} (u- \bar{u}_\text{b}) \right] \right\rbrace^2 \left( dr_\text{bp}^2 + r_\text{bp}^2 d\varphi^2 \right).
\end{align}
The approximation
\begin{align}\label{g_p_vr_abp_border}
& \left.  ds^2 \right|_{\mathcal{R}_\text{p} \cap \mathcal{D}_\text{a}} = - 2 du dv_\text{ap} + \left[ 1+ O \left( (u- \bar{u}_\text{a})^2 \right) \right] \left( dr_\text{ap}^2 + r_\text{ap}^2 d\varphi^2 \right), \nonumber \\
& \left.  ds^2 \right|_{\mathcal{R}_\text{p} \cap \mathcal{D}_\text{b}} = - 2 du dv_\text{bp}+ \left[ 1+ O \left( (u- \bar{u}_\text{b})^2 \right) \right] \left( dr_\text{bp}^2 + r_\text{bp}^2 d\varphi^2 \right).
\end{align}
will be used in Sec.\ \ref{Exterior_regions_Extending_the_Rosen_pulse_coordinates} to extend these coordinates beyond the respective boundaries $u=\bar{u}_\text{a}$ and $u=\bar{u}_\text{b}$ and connect the pulse region $\mathcal{R}_\text{p}$ with the exterior Minkowski spacetime.

\subsection{Gravitational wave}\label{Gravitational_wave}

The emission of the pulse is expected to generate a GW propagating in the same direction as the pulse and with the same planar wavefront. To explore this effect, we focus on the region $\mathcal{R}_\text{g}$, defined as the radial extension of $\mathcal{R}_\text{p}$ beyond its boundary at $r_{\text{p}\text{g}} = \bar{r}_\text{p}$. Similarly to Sec.\ \ref{Light_pulse}, we employ two coordinate systems in this region: one based on the assumed cylindrical symmetry of $\mathcal{R}_\text{g}$ and the other being locally Minkowskian near hyperplanes of constant $u$.

\subsubsection{Extending the Brinkmann-pulse coordinates to the GW region}

In Sec.\ \ref{Brinkmann_pulse_coordinates}, we introduced the Brinkmann coordinates $(u,v_{\text{p}\text{g}},r_{\text{p}\text{g}},\varphi)$ to describe the interior region of the pulse $\mathcal{R}_\text{p}$. A key advantage of this coordinate system is that it explicitly show the stationary and cylindrical symmetry of $\mathcal{R}_\text{p}$ through the coordinates $t_{\text{p}\text{g}} = (v_{\text{p}\text{g}} + u)/\sqrt{2}$, $z_{\text{p}\text{g}} = (v_{\text{p}\text{g}} - u)/\sqrt{2}$ and $\varphi$, along with the corresponding Killing vector fields $\partial_{t_{\text{p}\text{g}}}$, $\partial_{z_{\text{p}\text{g}}}$ and $\partial_{\varphi}$. These symmetries naturally lead to the use of $r_{\text{p}\text{g}}$ as the radial coordinate to define the boundary of the pulse at $r_{\text{p}\text{g}} = \bar{r}_\text{p}$, separating the regions $\mathcal{R}_\text{p}$ and $\mathcal{R}_\text{g}$. In this subsection, we extend the Brinkmann coordinate system $(u,v_{\text{p}\text{g}},r_{\text{p}\text{g}},\varphi)$ from $\mathcal{R}_\text{p}$ to $\mathcal{R}_\text{g}$ while preserving the symmetries.

We derive the metric in $\mathcal{R}_\text{g}$ by solving the vacuum Einstein equations while imposing stationary and cylindrical symmetries, and ensuring continuity conditions at the boundary $r_{\text{p}\text{g}} = \bar{r}_\text{p}$. We use the ansatz of Brinkmann coordinate system
\begin{equation}\label{g_g_vr_pg_r_Phi}
\left.  ds^2 \right|_{\mathcal{R}_\text{g}}  = - \Phi(r_{\text{p}\text{g}}) du^2 -2dudv_{\text{p}\text{g}} + dr_{\text{p}\text{g}}^2 +r_{\text{p}\text{g}}^2 d\varphi^2.
\end{equation}
This form ensures that $\partial_u$, $\partial_{v_{\text{p}\text{g}}}$ and $\partial_{\varphi}$ are Killing vector fields in $\mathcal{R}_\text{g}$. The metric in Eq.\ (\ref{g_g_vr_pg_r_Phi}) solves the vacuum Einstein equations $R_{\mu\nu} - R g_{\mu\nu}/2 = 0$ provided that $\Phi(r_{\text{p}\text{g}})$ satisfies the differential equation
\begin{equation}\label{Phi_g}
\Phi''(r_{\text{p} \text{g}})+\frac{\Phi'(r_{\text{p} \text{g}})}{r_{\text{p} \text{g}}} = 0.
\end{equation}

A unique solution to Eq.\ (\ref{Phi_g}) is determined by imposing continuity conditions for the metric at the boundary $r_{\text{p}\text{g}} = \bar{r}_\text{p}$ up to first-order derivatives. We allow discontinuities in second-order derivatives due to the stress-energy tensor, which is nonzero in $\mathcal{R}_\text{p}$ and zero in $\mathcal{R}_\text{g}$. Since the Einstein equations are second-order partial differential equations, these discontinuities manifest in the second derivatives of the metric at $r_{\text{p}\text{g}} = \bar{r}_\text{p}$. 

To impose the continuity condition, we consider the Taylor expansion of Eq.\ (\ref{g_p_vr_pg})
\begin{equation}\label{g_p_vr_pg_r_bar_p}
\left.  ds^2 \right|_{\mathcal{R}_\text{p}}  = -\epsilon \left[ 1 + \frac{2}{\bar{r}_\text{p}} \left( r_{\text{p}\text{g}} - \bar{r}_\text{p} \right)  + \mathcal{O} \left(\left(r_{\text{p}\text{g}} - \bar{r}_\text{p}\right)^2 \right) \right] du^2 -2dudv_{\text{p}\text{g}} + dr_{\text{p}\text{g}}^2 +r_{\text{p}\text{g}}^2 d\varphi^2,
\end{equation}
with $\epsilon = \kappa  \mathcal{E}  \bar{r}_\text{p}^2$. Requiring that the metric in $\mathcal{R}_\text{g}$ matches this expansion at $r_{\text{p}\text{g}} = \bar{r}_\text{p}$ we write
\begin{equation}\label{g_g_vr_pg_r_bar_p}
\left.  ds^2 \right|_{\mathcal{R}_\text{g}}  = -\epsilon \left[ 1 + \frac{2}{\bar{r}_\text{p}} \left( r_{\text{p}\text{g}} - \bar{r}_\text{p} \right)  + \mathcal{O} \left(\left(r_{\text{p}\text{g}} - \bar{r}_\text{p}\right)^2 \right) \right] du^2 -2dudv_{\text{p}\text{g}} + dr_{\text{p}\text{g}}^2 +r_{\text{p}\text{g}}^2 d\varphi^2 .
\end{equation}
By imposing Eq.\ (\ref{g_g_vr_pg_r_bar_p}), we obtain $\Phi(r_{\text{p}\text{g}}) = \epsilon \left[1 + 2\log(r_{\text{p}\text{g}}/\bar{r}_\text{p})  \right]$ as the unique solution for Eq.\ (\ref{Phi_g}). Thus, the metric in $\mathcal{R}_\text{g}$ takes the form
\begin{equation}\label{g_g_vr_pg_r}
\left.  ds^2 \right|_{\mathcal{R}_\text{g}}  = - \epsilon \left[1 + 2\log \left( \frac{r_{\text{p}\text{g}}}{\bar{r}_\text{p}} \right)  \right] du^2 -2dudv_{\text{p}\text{g}} + dr_{\text{p}\text{g}}^2 +r_{\text{p}\text{g}}^2 d\varphi^2.
\end{equation}

\subsubsection{Rosen-GW coordinates}

In Sec.\ \ref{Rosen_pulse_coordinates}, we introduced the Rosen-pulse coordinates $(u,v_{\xi\text{p}},r_{\xi\text{p}},\varphi)$ in $\mathcal{R}_\text{p}$. At $u = \bar{u}_\xi$, the associated metric (\ref{g_p_vr_xip}) can be locally approximated by the Minkowski metric. Here, we define the Rosen-GW coordinates $(u,v_{\xi\text{g}},r_{\xi\text{g}},\varphi)$ by requiring that the metric in $\mathcal{R}_\text{g}$ takes the form
\begin{equation}\label{g_g_vr_xig}
\left.  ds^2 \right|_{\mathcal{R}_\text{g}}  = - 2 du dv_{\xi\text{g}} + \left[ R_\text{g}'(u- \bar{u}_\xi, r_{\xi\text{g}}) \right]^2 dr_{\xi\text{g}}^2 +  R_\text{g}^2(u- \bar{u}_\xi, r_{\xi\text{g}}) d\varphi^2
\end{equation}
for some function $R_\text{g} (u , r )$ and that, at $u = \bar{u}_\xi$, the metric is locally Minkowskian:
\begin{equation}\label{g_g_vr_xig_u_xi}
\left.  ds^2 \right|_{\mathcal{R}_\text{g}} = - 2 du dv_{\xi\text{g}} + \left[ 1+ O \left( (u- \bar{u}_{\xi\text{g}})^2 \right) \right] \left( dr_{\xi\text{g}}^2 + r_{\xi\text{g}}^2 d\varphi^2 \right) .
\end{equation}

By using the ansatz 
\begin{equation}\label{vr_pg_vr_xig}
\left( u, v_{\text{p}\text{g}}, r_{\text{p}\text{g}}, \varphi \right) = \left( u, v_{\xi \text{g}} + V_\text{g} \left(u-\bar{u}_\xi ,r_{\xi \text{g}} \right), R_\text{g} \left(u-\bar{u}_\xi ,r_{\xi \text{g}} \right), \varphi \right),
\end{equation}
in Eq.\ (\ref{g_g_vr_pg_r}), we obtain the metric
\begin{align}\label{R_g_V_g_equations}
& \left.  ds^2 \right|_{\mathcal{R}_\text{g}}  = \left\lbrace - 2 \epsilon \log \left[\frac{R_{\text{g}}\left(u-\bar{u}_{\xi },r_{\xi \text{g}}\right)}{\bar{r}_{\text{p}}}\right]-\epsilon+\left[ \dot{R}_{\text{g}}\left(u-\bar{u}_{\xi },r_{\xi \text{g}}\right) \right]^2 -2 \dot{V}_{\text{g}}\left(u-\bar{u}_{\xi },r_{\xi \text{g}}\right) \right\rbrace du^2 - 2 du dv_{\xi \text{g}} \nonumber \\
&  +  2 \left[ \dot{R}_{\text{g}}\left(u-\bar{u}_{\xi },r_{\xi \text{g}}\right) R_{\text{g}}^{\prime }\left(u-\bar{u}_{\xi },r_{\xi \text{g}}\right)-V_{\text{g}}^{\prime }\left(u-\bar{u}_{\xi },r_{\xi \text{g}}\right) \right] du dr_{\xi \text{g}} +\left[ R_{\text{g}}^{\prime }\left(u-\bar{u}_{\xi },r_{\xi \text{g}}\right)\right]^2 dr_{\xi \text{g}}^2 + R_{\text{g}}^2\left(u-\bar{u}_{\xi },r_{\xi \text{g}}\right) d \varphi^2.
\end{align}
To match this solution with Eq.\ (\ref{g_g_vr_xig}), the following differential equations must be satisfied:
\begin{subequations}\label{R_g_V_g_equations_2}
\begin{align}
&  - 2 \epsilon \log \left[\frac{R_{\text{g}}\left(u-\bar{u}_{\xi },r_{\xi \text{g}}\right)}{\bar{r}_{\text{p}}}\right]-\epsilon+\left[\dot{R}_{\text{g}}\left(u-\bar{u}_{\xi },r_{\xi \text{g}}\right)\right]^2-2 \dot{V}_{\text{g}}\left(u-\bar{u}_{\xi },r_{\xi \text{g}}\right) = 0, \label{R_g_V_g_equations_a} \\
& \dot{R}_{\text{g}}\left(u-\bar{u}_{\xi },r_{\xi \text{g}}\right) R_{\text{g}}^{\prime }\left(u-\bar{u}_{\xi },r_{\xi \text{g}}\right)-V_{\text{g}}^{\prime }\left(u-\bar{u}_{\xi },r_{\xi \text{g}}\right)  = 0. \label{R_g_V_g_equations_b} 
\end{align}
\end{subequations}
The boundary conditions are determined by matching Eq.\ (\ref{R_g_V_g_equations}) with Eq.\ (\ref{g_g_vr_xig_u_xi}), resulting in $ R_{\text{g}}\left(0,r_{\xi \text{g}}\right) = 0$ and $ R_{\text{g}}^{\prime }\left(0,r_{\xi \text{g}}\right) = 0 $.

By differentiating Eq.\ (\ref{R_g_V_g_equations_a}) with respect to the variable $r_{\xi \text{g}}$ and Eq.\ (\ref{R_g_V_g_equations_b}) with respect to $u$, the two equations decouple and give a single differential equation for $R_{\text{g}}\left(u-\bar{u}_{\xi },r_{\xi \text{g}}\right)$:
\begin{equation}
\ddot{R}_{\text{g}}\left(u-\bar{u}_{\xi },r_{\xi \text{g}}\right) R_{\text{g}}\left(u-\bar{u}_{\xi },r_{\xi \text{g}}\right) + \epsilon = 0.
\end{equation}
The solution to this equation, subject to the boundary conditions $ R_{\text{g}}\left(0,r_{\xi \text{g}}\right) = 0$ and $ R_{\text{g}}^{\prime }\left(0,r_{\xi \text{g}}\right) = 0 $, is given by 
\begin{equation} \label{R_g}
R_\text{g} (u , r ) = \exp \left\lbrace - \left[ \text{erf}^{-1} \left( \sqrt{\frac{2 \epsilon }{\pi }} \frac{u}{r}\right) \right]^2\right\rbrace r.
\end{equation}
By substituting Eq.\ (\ref{R_g}) into Eq.\ (\ref{R_g_V_g_equations_2}), we finally obtain
\begin{equation}\label{V_g}
V_{\text{g}} (u,r) = -\sqrt{2 \epsilon } \, \text{erf}^{-1} \left( \sqrt{\frac{2 \epsilon }{\pi }} \frac{u}{r}\right)\exp \left\lbrace - \left[ \text{erf}^{-1} \left( \sqrt{\frac{2 \epsilon }{\pi }} \frac{u}{r}\right) \right]^2\right\rbrace r +  \frac{\epsilon}{2} \left[1-2 \log \left(\frac{r}{\bar{r}_{\text{p}}}\right)\right] u.
\end{equation}

\subsection{Exterior regions}\label{Exterior_regions}

The pulse and the GW are confined within $\bar{u}_\text{a} \leq u \leq \bar{u}_\text{b}$.  In this section, we extend our analysis beyond these boundaries, examining the two exterior regions $\mathcal{R}_\text{a}$ and $\mathcal{R}_\text{b}$, respectively defined by $u < \bar{u}_\text{a}$ and $u > \bar{u}_\text{b}$.

\subsubsection{Extending the Rosen-pulse coordinates}\label{Exterior_regions_Extending_the_Rosen_pulse_coordinates}

In this subsection, we extend the Rosen-pulse coordinates $(u,v_{\text{ap}},r_{\text{ap}},\varphi)$ and $(u,v_{\text{bp}},r_{\text{bp}},\varphi)$, introduced in Sec.\ \ref{Rosen_pulse_coordinates}, to the exterior regions $\mathcal{R}_\text{a}$ and $\mathcal{R}_\text{b}$, respectively. Near the respective boundaries $u = \bar{u}_\text{a}$ and $u = \bar{u}_\text{b}$, the metric given by Eq.\ (\ref{g_p_vr_abp}) can be approximated as in Eq.\ (\ref{g_p_vr_abp_border}). To satisfy continuity conditions, the metric in the exterior regions $\mathcal{R}_\text{a}$ and $\mathcal{R}_\text{b}$, must approximate as
\begin{align}\label{g_ab_vr_abp_border}
& \left.  ds^2 \right|_{\mathcal{R}_\text{a}} = - 2 du dv_\text{ap} + \left[ 1+ O \left( (u- \bar{u}_\text{a})^2 \right) \right] \left( dr_\text{ap}^2 + r_\text{ap}^2 d\varphi^2 \right), \nonumber \\
& \left.  ds^2 \right|_{\mathcal{R}_\text{b}} = - 2 du dv_\text{bp}+ \left[ 1+ O \left( (u- \bar{u}_\text{b})^2 \right) \right] \left( dr_\text{bp}^2 + r_\text{bp}^2 d\varphi^2 \right).
\end{align}
Discontinuities in the second-order derivatives arise from the stress-energy tensor, which is uniformly nonzero in $\mathcal{R}_\text{p}$ and identically zero in $\mathcal{R}_\text{a}$ and $\mathcal{R}_\text{b}$.

In addition to the boundary conditions (\ref{g_ab_vr_abp_border}) at $u = \bar{u}_\text{a}$ and $u = \bar{u}_\text{b}$, we impose initial conditions on the metric $g_{\mu\nu}$ in $\mathcal{R}_\text{a}$ and $\mathcal{R}_\text{b}$ by assuming that, at past null infinity (i.e., $u \to -\infty$ and $v_{\xi\text{p}} \to -\infty$), $g_{\mu\nu}$ reduces to the Minkowski metric. Also, we assume that the stress-energy tensor is zero throughout the entire exterior region $\mathcal{R}_\text{a} \cup \mathcal{R}_\text{b}$. Under these conditions, we find that in the regions $\mathcal{R}_\text{a}$ and $\mathcal{R}_\text{b}$, the respective coordinates $(u,v_{\text{ap}},r_{\text{ap}},\varphi)$ and $(u,v_{\text{bp}},r_{\text{bp}},\varphi)$ describe a Minkowski spacetime:
\begin{align}\label{g_ab_vr_abp}
& \left.  ds^2 \right|_{\mathcal{R}_\text{a}} = - 2 du dv_\text{ap} + dr_\text{ap}^2 + r_\text{ap}^2 d\varphi^2, & \left.  ds^2 \right|_{\mathcal{R}_\text{b}} = - 2 du dv_\text{bp}+ dr_\text{bp}^2 + r_\text{bp}^2 d\varphi^2.
\end{align}

\subsubsection{Extending the Rosen-GW coordinates}

To extend the Rosen-GW coordinates $(u,v_{\text{ag}},r_{\text{ag}},\varphi)$ and $(u,v_{\text{bg}},r_{\text{bg}},\varphi)$ into $\mathcal{R}_\text{a}$ and $\mathcal{R}_\text{b}$, respectively, we follow the same approach as in Sec.\ \ref{Exterior_regions_Extending_the_Rosen_pulse_coordinates}. We use Eq.\ (\ref{g_g_vr_xig_u_xi}) for \R{any} $\xi \R{\in} \{ \text{a}, \text{b} \}$ and impose the continuity condition at $u=\bar{u}_\xi$, which leads to
\begin{align}\label{g_ab_vr_abg_border}
& \left.  ds^2 \right|_{\mathcal{R}_\text{a}} = - 2 du dv_\text{ag} + \left[ 1+ O \left( (u- \bar{u}_\text{a})^2 \right) \right] \left( dr_\text{ag}^2 + r_\text{ag}^2 d\varphi^2 \right), \nonumber \\
& \left.  ds^2 \right|_{\mathcal{R}_\text{b}} = - 2 du dv_\text{bg}+ \left[ 1+ O \left( (u- \bar{u}_\text{b})^2 \right) \right] \left( dr_\text{bg}^2 + r_\text{bg}^2 d\varphi^2 \right).
\end{align}
Additionally, we assume that at past null infinity of $\mathcal{R}_\text{a}$ and $\mathcal{R}_\text{b}$, the coordinates  $(u,v_{\text{ag}},r_{\text{ag}},\varphi)$ and $(u,v_{\text{bg}},r_{\text{bg}},\varphi)$ describe a Minkowski spacetime. As a result, we obtain
\begin{align}\label{g_ab_vr_abg}
& \left.  ds^2 \right|_{\mathcal{R}_\text{a}} = - 2 du dv_\text{ag} +  dr_\text{ag}^2 + r_\text{ag}^2 d\varphi^2, & \left.  ds^2 \right|_{\mathcal{R}_\text{b}} = - 2 du dv_\text{bg}+ dr_\text{bg}^2 + r_\text{bg}^2 d\varphi^2.
\end{align}

In each region $\xi \R{\in} \{ \text{a}, \text{b} \}$, both the Rosen-pulse coordinates $(u,v_{\xi\text{p}},r_{\xi\text{p}},\varphi)$ and the Rosen-GW coordinates $(u,v_{\xi\text{g}},r_{\xi\text{g}},\varphi)$ describe a Minkowski spacetime. Hence, there must exist a Poincaré transformation mapping $(u,v_{\xi\text{p}},r_{\xi\text{p}},\varphi) \mapsto (u,v_{\xi\text{g}},r_{\xi\text{g}},\varphi)$, such that
\begin{equation}
\begin{pmatrix}
(v_{\xi\text{g}} + u)/\sqrt{2} \\
r_{\xi\text{g}} \cos(\varphi)\\
r_{\xi\text{g}} \sin(\varphi) \\
(v_{\xi\text{g}} - u)/\sqrt{2}
\end{pmatrix}
= \Lambda 
\begin{pmatrix}
(v_{\xi\text{p}} + u)/\sqrt{2} \\
r_{\xi\text{p}} \cos(\varphi)\\
r_{\xi\text{p}} \sin(\varphi) \\
(v_{\xi\text{p}} - u)/\sqrt{2}
\end{pmatrix} + \alpha,
\end{equation}
where $\Lambda$ is a Lorentz matrix and $\alpha$ is a four-vector representing a translation. However, such a transformation reduces to the identity map $(u,v_{\xi\text{p}},r_{\xi\text{p}},\varphi) = (u,v_{\xi\text{g}},r_{\xi\text{g}},\varphi)$. This can be verified by noticing that the transformations $(u,v_{\text{p}\text{g}},r_{\text{p}\text{g}},\varphi) \mapsto (u,v_{\xi\text{p}},r_{\xi\text{p}},\varphi)$ [Eq.\ (\ref{vr_pg_vr_xip})] and $(u,v_{\text{p}\text{g}},r_{\text{p}\text{g}},\varphi) \mapsto (u,v_{\xi\text{g}},r_{\xi\text{g}},\varphi)$ [Eq.\ (\ref{vr_pg_vr_xig})] reduce to the identity when $u = \bar{u}_\xi$. Hence, in the limit $u \to \bar{u}_\xi$ and $r_\text{pg} \to \bar{r}_\text{p}$, the two coordinate systems $(u,v_{\xi\text{p}},r_{\xi\text{p}},\varphi)$ and $(u,v_{\xi\text{g}},r_{\xi\text{g}},\varphi)$ coincide. The only Poincaré transformation 
$(\Lambda, \alpha)$ satisfying
\begin{equation}
\begin{pmatrix}
(v_{\xi\text{p}} + \bar{u}_\xi)/\sqrt{2} \\
\bar{r}_\text{p} \cos(\varphi)\\
\bar{r}_\text{p} \sin(\varphi) \\
(v_{\xi\text{p}} - \bar{u}_\xi)/\sqrt{2}
\end{pmatrix}
= \Lambda 
\begin{pmatrix}
(v_{\xi\text{p}} + \bar{u}_\xi)/\sqrt{2} \\
\bar{r}_\text{p} \cos(\varphi)\\
\bar{r}_\text{p} \sin(\varphi) \\
(v_{\xi\text{p}} - \bar{u}_\xi)/\sqrt{2}
\end{pmatrix} + \alpha
\end{equation}
for any $v_{\xi\text{p}}$ and $\varphi$ is the identity map $(\Lambda, \alpha) = (\mathbb{I}, 0)$.

Since the two coordinate systems $(u,v_{\xi\text{p}},r_{\xi\text{p}},\varphi)$ and $(u,v_{\xi\text{g}},r_{\xi\text{g}},\varphi)$ are identical in each region $\mathcal{R}_\xi$, we define a unified coordinate system $(u,v,r,\varphi)$ such that
\begin{equation}\label{vr_vr_xip_vr_xig}
\left. (u,v,r,\varphi) \right|_{\mathcal{R}_\xi} = \left. (u,v_{\xi\text{p}},r_{\xi\text{p}},\varphi) \right|_{\mathcal{R}_\xi} = \left. (u,v_{\xi\text{g}},r_{\xi\text{g}},\varphi) \right|_{\mathcal{R}_\xi}.
\end{equation}
These coordinates describe the exterior region $\mathcal{R}_\text{a} \cup \mathcal{R}_\text{b}$ as a Minkowski spacetime
\begin{equation}
\left.  ds^2 \right|_{\mathcal{R}_\text{a} \cup \mathcal{R}_\text{b}} = - 2 du dv +  dr^2 + r^2 d\varphi^2 .
\end{equation}

\section{Particle trajectories}

In Sec.\ \ref{Spacetime_of_a_single_pulse}, we described the metric associated with a single pulse. In this section, we use that metric to derive the corresponding geodesics. Our focus is on particles traversing the GW region $\mathcal{R}_\text{g}$. We show how the trajectory of particles is perturbed by the passage of the GW.

\subsection{Particles propagating inside the GW}\label{Brinkmann_pulse_coordinates_Geodesics}

In this subsection, we derive the general trajectory of a particle traveling within $\mathcal{R}_\text{g}$ with a constant angular coordinate $\varphi = 0$. By using Eq.\ (\ref{g_p_vr_pg}), we solve the geodesic equation 
\begin{equation}\label{geodesic_equation}
\frac{d^2 x^\rho}{ds^2}+ \Gamma^\rho{}_{\mu\nu} \frac{d x^\mu}{ds} \frac{d x^\nu}{ds}  = 0
\end{equation}
for a particle moving in $\mathcal{R}_\text{g}$. The trajectory is described by using the coordinate system $(u, v_{\text{p}\text{g}}, r_{\text{p}\text{g}}, \varphi)$ and parameterized by $s$ as $ (x^0(s), x^1(s), x^2(s), x^3(s)) = (u(s), v_{\text{p}\text{g}}(s), r_{\text{p}\text{g}}(s), \varphi(s))$. The Christoffel symbols $\Gamma^\rho{}_{\mu\nu} $ are determined from the metric (\ref{g_p_vr_pg}).

By imposing the condition $\varphi(s) = 0$, the geodesic equations (\ref{geodesic_equation}) reduce to
\begin{align}\label{geodesic_equation_g_pg}
& u''(s) = 0, && v_{\text{pg}}''(s) = - 2 \epsilon  \frac{u'(s) r_{\text{pg}}'(s)}{r_{\text{pg}}(s)} , && r_{\text{pg}}''(s) =  -\epsilon \frac{ \left[u'(s)\right]^2}{r_{\text{pg}}(s)}.
\end{align}
The first equation in (\ref{geodesic_equation_g_pg}) leads to two cases: $u(s)$ being constant ($u(s) = \bar{u}_\gamma$) and $u(s)$ being linear in $s$. In the constant case, the four-velocity $U^\mu = dx^\mu/ds$ satisfies $ U^\mu(s) U_\mu(s) = [r_{\text{p}\text{g} }'(s)]^2$. The condition of no superluminality $U^\mu(s) U_\mu(s) \leq 0$ implies that $r_{\text{p}\text{g}}(s) $ must be constant (denoted as $r_{\text{p}\text{g}}(s) = \bar{r}_\gamma$) and the particle must be massless. For constant $u(s)$, the second equation in (\ref{geodesic_equation_g_pg}) simplifies to $v_{\text{p} \text{g}}''(s) = 0$, which implies that $v_{\text{p} \text{g}}(s) $ is linear in $s$. By choosing a convenient affine parametrization, we can write $v_{\text{p} \text{g}}(s) = s$. Hence, the general solution to Eq.\ (\ref{geodesic_equation_g_pg}) for constant $u(s)$ is
\begin{equation}\label{geodesic_u_gamma}
\left.(u(s), v_{\text{p}\text{g}}(s), r_{\text{p}\text{g}}(s), \varphi(s)) \right|_{\mathcal{R}_\text{g}} = (\bar{u}_\gamma, s, \bar{r}_\gamma, 0),
\end{equation}
which describes the trajectory of a massless particle traveling in the same direction as the pulse.

If $u(s)$ is not constant, the first equation in (\ref{geodesic_equation_g_pg}) implies that $u(s)$ must be linear in $s$. We can translate $s$ to write $u(s) = c_{\text{g},1} s$, where $c_{\text{g},1}$ is a nonzero constant. Under this assumption, the second and third equations simplify to
\begin{align}\label{geodesic_equation_g_pg_3}
& v_{\text{pg}}''(s) =  - 2 \epsilon  c_{\text{g},1} \frac{ r_{\text{pg}}'(s)}{r_{\text{pg}}(s)}, & r_{\text{pg}}''(s) = - \frac{\epsilon  c_{\text{g},1}^2}{r_{\text{pg}}(s)}.
\end{align}
The solutions to these equations are given by
\begin{align}
& v_{\text{pg}}(s) = c_{\text{g},2} s + c_{\text{g},3} +V_\text{g} \left( c_{\text{g},1} s + c_{\text{g},4}, c_{\text{g},5} \right), & r_{\text{pg}}(s)  = R_\text{g} \left( c_{\text{g},1} s + c_{\text{g},4}, c_{\text{g},5}\right),
\end{align}
where the functions $V_{\text{g}} (u,r)$ and $R_{\text{g}} (u,r)$ are defined in Eqs.\ (\ref{V_g}) and (\ref{R_g}). Thus, the general solution to Eq.\ (\ref{geodesic_equation_g_pg}) for nonconstant $u(s)$ and constant $\varphi(s) = 0$ is
\begin{equation}\label{geodesics_g}
\left. (u(s), v_{\text{p}\text{g}}(s), r_{\text{p}\text{g}}(s), \varphi(s)) \right|_{\mathcal{R}_\text{g}}  = (c_{\text{g},1} s, c_{\text{g},2} s + c_{\text{g},3} +V_\text{g} \left( c_{\text{g},1} s + c_{\text{g},4}, c_{\text{g},5} \right), R_\text{g} \left( c_{\text{g},1} s + c_{\text{g},4}, c_{\text{g},5}\right), 0).
\end{equation}
The corresponding four-velocity is
\begin{align}\label{velocity_g}
\left. U(s) \right|_{\mathcal{R}_\text{g}} = & c_{\text{g},1} \partial_u + \left\lbrace c_{\text{g},2} -\epsilon  c_{\text{g},1} + 2 \epsilon  c_{\text{g},1} \left[ \text{erf}^{-1}\left( \sqrt{\frac{2 \epsilon}{\pi }} \frac{c_{\text{g},1} s + c_{\text{g},4} }{c_{\text{g},5}} \right) \right]^2   + \frac{\epsilon }{2}  c_{\text{g},1} \left[1-2 \log \left(\frac{c_{\text{g},5}}{\bar{r}_{\text{p}}}\right)\right] \right\rbrace \partial_{v_{\text{p}\text{g}}} \nonumber \\
& -\sqrt{2 \epsilon } c_{\text{g},1} \text{erf}^{-1}\left( \sqrt{\frac{2 \epsilon}{\pi }} \frac{c_{\text{g},1} s + c_{\text{g},4} }{c_{\text{g},5}} \right)  \partial_{r_{\text{p}\text{g}}}.
\end{align}

\subsection{Particle-GW interaction}\label{Particle_GW_interaction}

In this subsection, we assume that the particle, initially traveling in the Minkowski region $\mathcal{R}_\text{a}$, crosses the boundary between $\mathcal{R}_\text{a}$ and $\mathcal{R}_\text{g}$. Then, it exits $\mathcal{R}_\text{g}$ and enters the other Minkowski region $\mathcal{R}_\text{b}$, where its trajectory will differ from the initial one. Throughout its motion, the particle never enters the region $\mathcal{R}_\text{p}$. 

In the Minkowski spacetime, geodesics are straight lines. In particular, particles traveling in the semiplane $\varphi = 0$ are described by
\begin{align}\label{geodesics_xig}
& \left. (u(s),v(s),r(s),\varphi(s)) \right|_{\mathcal{R}_\text{a}} = \left( c_{\text{a},1} s , c_{\text{a},2} s + c_{\text{a},3},  c_{\text{a},4} s + c_{\text{a},5} ,  0 \right), \nonumber \\
& \left. (u(s),v(s),r(s),\varphi(s)) \right|_{\mathcal{R}_\text{b}} = \left( c_{\text{b},1} s , c_{\text{b},2} s + c_{\text{b},3},  c_{\text{b},4} s + c_{\text{b},5} ,  0 \right).
\end{align}
To extend these geodesics into $\mathcal{R}_\text{g} $ across the boundaries $u = \bar{u}_\text{a}$ and $u = \bar{u}_\text{b}$, we use the coordinate transformations (\ref{vr_pg_vr_xig}) and (\ref{vr_vr_xip_vr_xig}) and impose the continuity conditions $ \left. (u(s),v_{\xi\text{g}}(s),r_{\xi\text{g}}(s),\varphi(s)) \right|_{\mathcal{R}_\text{g}} = \left. (u(s),v_{\xi\text{g}}(s),r_{\xi\text{g}}(s),\varphi(s)) \right|_{\mathcal{R}_\xi}$ and $\left. U(s) \right|_{\mathcal{R}_\text{g}} = \left. U(s) \right|_{\mathcal{R}_\xi}$ at $u = \bar{u}_\xi$ for \R{any} $\xi \R{\in} \{ \text{a}, \text{b} \}$, which lead to
\begin{align}\label{c_g_c_xi}
& \begin{aligned} & c_{\text{g},1} =  c_{\xi ,1}, & c_{\text{g},2} =  c_{\xi ,2}-\frac{c_{\xi ,4}^2}{2 c_{\xi ,1}}, \end{aligned} \nonumber \\
& \begin{aligned} c_{\text{g},3} = & c_{\xi ,3} -\frac{c_{\xi ,4}}{c_{\xi ,1}}  \left(   c_{\xi ,5} + \frac{ c_{\xi ,4}}{2 c_{\xi ,1}} \bar{u}_{\xi } \right) \nonumber \\
&   -  \frac{\sqrt{\pi}}{2 \sqrt{2 \epsilon }}\exp \left( \frac{c_{\xi ,4}^2}{2 \epsilon  c_{\xi ,1}^2} \right) \text{erf}\left(\frac{c_{\xi ,4}}{\sqrt{2 \epsilon } c_{\xi ,1}}\right) \left[\frac{c_{\xi ,4}^2}{ c_{\xi ,1}^2}-\epsilon + 2 \epsilon \log \left( \frac{c_{\xi ,1} c_{\xi ,5} + c_{\xi ,4}\bar{u}_{\xi }}{c_{\xi ,1} \bar{r}_{\text{p}} }\right) \right] \left(  c_{\xi ,5} + \frac{ c_{\xi ,4}}{ c_{\xi ,1}} \bar{u}_{\xi } \right), \end{aligned} \nonumber \\
& \begin{aligned}  & c_{\text{g},4} = -\bar{u}_{\xi }  - \sqrt{\frac{\pi }{2\epsilon}} \exp \left( \frac{c_{\xi ,4}^2}{2 \epsilon  c_{\xi ,1}^2} \right)   \text{erf}\left(\frac{c_{\xi ,4}}{\sqrt{2\epsilon } c_{\xi ,1}}\right) \left(   c_{\xi ,5} + \frac{ c_{\xi ,4}}{ c_{\xi ,1}} \bar{u}_{\xi } \right), & c_{\text{g},5} =  \exp \left( \frac{c_{\xi ,4}^2}{2 \epsilon  c_{\xi ,1}^2} \right)  \left(  c_{\xi ,5} + \frac{ c_{\xi ,4}}{ c_{\xi ,1}} \bar{u}_{\xi } \right) . \end{aligned}
\end{align}
The inverse map $(c_{\text{g},1}, \dots, c_{\text{g},5}) \mapsto (c_{\xi ,1}, \dots, c_{\xi ,5})$ is
\begin{align}\label{c_xi_c_g}
& \begin{aligned} & c_{\xi ,1} =  c_{\text{g},1}, & c_{\xi ,2} = c_{\text{g},2} + \epsilon  c_{\text{g},1} \left[ \text{erf}^{-1}\left( \sqrt{\frac{2 \epsilon}{\pi }}  \frac{c_{\text{g},4} + \bar{u}_{\xi }}{c_{\text{g},5}}\right) \right]^2, \end{aligned} \nonumber \\
&  \begin{aligned} c_{\xi ,3} = & c_{\text{g},3} + \frac{\epsilon}{2} \left( c_{\text{g},4} + \bar{u}_{\xi } \right) -\sqrt{2 \epsilon } \exp \left\lbrace - \left[ \text{erf}^{-1}\left( \sqrt{\frac{2 \epsilon}{\pi }}  \frac{c_{\text{g},4} + \bar{u}_{\xi }}{c_{\text{g},5}}\right) \right]^2\right\rbrace \text{erf}^{-1}\left( \sqrt{\frac{2 \epsilon}{\pi }}  \frac{c_{\text{g},4} + \bar{u}_{\xi }}{c_{\text{g},5}}\right) c_{\text{g},5}  \nonumber \\
& - \epsilon \left[  \text{erf}^{-1}\left( \sqrt{\frac{2 \epsilon}{\pi }}  \frac{c_{\text{g},4} + \bar{u}_{\xi }}{c_{\text{g},5}}\right) \right]^2  \bar{u}_{\xi } - \epsilon  \log \left(\frac{c_{\text{g},5}}{\bar{r}_{\text{p}}}\right)  \left(c_{\text{g},4} + \bar{u}_{\xi }\right), \end{aligned}\nonumber \\
& c_{\xi ,4} =  - \sqrt{2 \epsilon } c_{\text{g},1}  \text{erf}^{-1}\left( \sqrt{\frac{2 \epsilon}{\pi }}  \frac{c_{\text{g},4} + \bar{u}_{\xi }}{c_{\text{g},5}}\right), \nonumber \\
&  c_{\xi ,5} = \exp \left\lbrace - \left[ \text{erf}^{-1}\left( \sqrt{\frac{2 \epsilon}{\pi }}  \frac{c_{\text{g},4} + \bar{u}_{\xi }}{c_{\text{g},5}}\right) \right]^2\right\rbrace c_{\text{g},5}  +  \sqrt{2 \epsilon }  \text{erf}^{-1}\left( \sqrt{\frac{2 \epsilon}{\pi }}  \frac{c_{\text{g},4} + \bar{u}_{\xi }}{c_{\text{g},5}}\right) \bar{u}_{\xi }.
\end{align}

We assume that the event in which the particle is hit by the GW is given by $(u, v, r, \varphi) = (\bar{u}_{\text{a}}, \bar{v}_{\text{a}}, \bar{r}_{\text{a}}, 0)$. At that moment, the particle has the four-velocity $U = U_{\text{a}}^u \partial_u + U_{\text{a}}^v \partial_v + U_{\text{a}}^r \partial_r$. Consequently, the parameters $(c_{\text{a},1}, \dots, c_{\text{a},5})$ are given by
\begin{align}\label{c_a_initial}
& c_{\text{a},1} = U_{\text{a}}^u, && c_{\text{a},2} = U_{\text{a}}^v, && c_{\text{a},3} = \bar{v}_{\text{a}}-\frac{U_{\text{a}}^v}{U_{\text{a}}^u} \bar{u}_{\text{a}}, && c_{\text{a},4} =  U_{\text{a}}^r, && c_{\text{a},5} =  \bar{r}_{\text{a}}-\frac{U_{\text{a}}^r}{U_{\text{a}}^u} \bar{u}_{\text{a}} .
\end{align}
After being hit by the GW, the particle will be found in $(u,v,r,\varphi) = (\bar{u}_\text{b},\bar{v}_\text{b},\bar{r}_\text{b},0)$ with four-velocity $U = U_{\text{b}}^u \partial_u + U_{\text{b}}^v \partial_v + U_{\text{b}}^r \partial_r$. By using Eq.\ (\ref{geodesics_xig}) with $\xi = \text{b}$, we can express $(\bar{v}_\text{b},\bar{r}_\text{b},U_{\text{b}}^u, U_{\text{b}}^v , U_{\text{b}}^r)$ in terms of the parameters $(c_{\text{b},1}, \dots c_{\text{b},5})$, resulting in
\begin{align}\label{final_c_b}
& \bar{v}_\text{b} = c_{\text{b},3} + \frac{c_{\text{b},2}}{c_{\text{b},1}} \bar{u}_{\text{b}} , && \bar{r}_\text{b} =  c_{\text{b},5} + \frac{ c_{\text{b},4}}{c_{\text{b},1}} \bar{u}_{\text{b}} , && U_{\text{b}}^u = c_{\text{b},1},  && U_{\text{b}}^v = c_{\text{b},2}, && U_{\text{b}}^r = c_{\text{b},4}.
\end{align}

To express the final parameters $(\bar{v}_\text{b},\bar{r}_\text{b},U_{\text{b}}^u, U_{\text{b}}^v , U_{\text{b}}^r)$ in terms of the initial data  $(\bar{v}_\text{a},\bar{r}_\text{a},U_{\text{a}}^u, U_{\text{a}}^v , U_{\text{a}}^r)$, we substitute Eq.\ (\ref{c_a_initial}) into Eq.\ (\ref{c_g_c_xi}) for $\xi=\text{a}$ and use the resulting equation in Eq.\ (\ref{c_xi_c_g}) for $\xi=\text{b}$. Finally, we apply this result to Eq.\ (\ref{final_c_b}) and obtain
\begin{align}\label{final_GW_initial}
& \begin{aligned} \bar{v}_\text{b} = \, & \bar{v}_{\text{a}}  - \frac{U_{\text{a}}^r}{U_{\text{a}}^u} \bar{r}_{\text{a}}  + \left[ \frac{U_{\text{a}}^v}{U_{\text{a}}^u} - \left(\frac{U_{\text{a}}^r}{U_{\text{a}}^u}\right)^2 +  \frac{\epsilon}{2} -  \epsilon \log \left(\frac{\bar{r}_{\text{a}}}{\bar{r}_{\text{p}}} \right) \right] \left( \bar{u}_{\text{b}} -  \bar{u}_{\text{a}} \right) \nonumber \\
 & + \sqrt{2 \epsilon } \, \text{erf}^{-1}\left\lbrace - \sqrt{\frac{2 \epsilon}{\pi}}  \exp\left[ - \frac{1}{2 \epsilon}  \left( \frac{U_{\text{a}}^r}{ U_{\text{a}}^u} \right)^2  \right]   \frac{\bar{u}_{\text{b}}-\bar{u}_{\text{a}}}{\bar{r}_{\text{a}}} + \text{erf}\left( \sqrt{\frac{1}{2 \epsilon} }  \frac{U_{\text{a}}^r}{U_{\text{a}}^u}\right)\right\rbrace \nonumber \\
 & \times \exp \left[ \frac{1}{2 \epsilon  } \left( \frac{U_{\text{a}}^r}{U_{\text{a}}^u} \right)^2 - \left( \text{erf}^{-1}\left\lbrace - \sqrt{\frac{2 \epsilon}{\pi}}  \exp\left[ - \frac{1}{2 \epsilon}  \left( \frac{U_{\text{a}}^r}{ U_{\text{a}}^u} \right)^2  \right]   \frac{\bar{u}_{\text{b}}-\bar{u}_{\text{a}}}{\bar{r}_{\text{a}}} + \text{erf}\left( \sqrt{\frac{1}{2 \epsilon} }  \frac{U_{\text{a}}^r}{U_{\text{a}}^u}\right)\right\rbrace \right)^2 \right] \bar{r}_{\text{a}}, \end{aligned}  \nonumber \\
& \bar{r}_\text{b} =  \exp \left[ \frac{1}{2 \epsilon  } \left( \frac{U_{\text{a}}^r}{U_{\text{a}}^u} \right)^2 - \left( \text{erf}^{-1}\left\lbrace - \sqrt{\frac{2 \epsilon}{\pi}}  \exp\left[ - \frac{1}{2 \epsilon}  \left( \frac{U_{\text{a}}^r}{ U_{\text{a}}^u} \right)^2  \right]   \frac{\bar{u}_{\text{b}}-\bar{u}_{\text{a}}}{\bar{r}_{\text{a}}} + \text{erf}\left( \sqrt{\frac{1}{2 \epsilon} }  \frac{U_{\text{a}}^r}{U_{\text{a}}^u}\right)\right\rbrace \right)^2 \right] \bar{r}_{\text{a}}, \nonumber \\
& \begin{aligned}  & U_{\text{b}}^u = U_{\text{a}}^u, & U_{\text{b}}^v = U_{\text{a}}^v -\frac{\left(U_{\text{a}}^r\right)^2}{2 U_{\text{a}}^u}+ \epsilon  \left( \text{erf}^{-1}\left\lbrace - \sqrt{\frac{2 \epsilon}{\pi}}  \exp\left[ - \frac{1}{2 \epsilon}  \left( \frac{U_{\text{a}}^r}{ U_{\text{a}}^u} \right)^2  \right]   \frac{\bar{u}_{\text{b}}-\bar{u}_{\text{a}}}{\bar{r}_{\text{a}}} + \text{erf}\left( \sqrt{\frac{1}{2 \epsilon} }  \frac{U_{\text{a}}^r}{U_{\text{a}}^u}\right)\right\rbrace \right)^2 U_{\text{a}}^u , \end{aligned}  \nonumber \\
& U_{\text{b}}^r = \sqrt{2 \epsilon }  \text{erf}^{-1}\left\lbrace - \sqrt{\frac{2 \epsilon}{\pi}}  \exp\left[ - \frac{1}{2 \epsilon}  \left( \frac{U_{\text{a}}^r}{ U_{\text{a}}^u} \right)^2  \right]   \frac{\bar{u}_{\text{b}}-\bar{u}_{\text{a}}}{\bar{r}_{\text{a}}} + \text{erf}\left( \sqrt{\frac{1}{2 \epsilon} }  \frac{U_{\text{a}}^r}{U_{\text{a}}^u}\right)\right\rbrace U_{\text{a}}^u.
\end{align}
\R{The limit $\bar{r}_\text{a} \to \infty$ results in $U_{\text{b}}^u \approx U_{\text{a}}^u$, $ U_{\text{b}}^v \approx U_{\text{a}}^v $ and $ U_{\text{b}}^r \approx U_{\text{a}}^r$, as stated in the main paper. Also,} the limit $\epsilon \to 0$ of Eq.\ (\ref{final_GW_initial}), evaluated at the first nontrivial order \R{and under the assumption of $U_\text{a}^u = U_\text{a}^v = 1/\sqrt{2}$ and $U_\text{a}^r=0$ leads to} Eq.\ (10) of the main paper.

After interacting with the GW, the particle continues to move through the Minkowski spacetime $\mathcal{R}_\text{b}$ at a constant velocity. At any $u = \bar{u}_\text{c} > \bar{u}_\text{b}$ and before crossing the center at $r=0$, the particle will be found in $(u, v, r, \varphi) = (\bar{u}_\text{c}, \bar{v}_\text{c}, \bar{r}_\text{c}, 0)$ with four-velocity $U = U_{\text{c}}^u \partial_u + U_{\text{c}}^v \partial_v + U_{\text{c}}^r \partial_r$ given by
\begin{align}\label{gamma_c}
& \begin{aligned} \bar{v}_\text{c} = \, & \bar{v}_{\text{a}}  - \frac{U_{\text{a}}^r}{U_{\text{a}}^u} \bar{r}_{\text{a}}  + \frac{U_{\text{a}}^v}{U_{\text{a}}^u} \left( \bar{u}_{\text{c}} - \bar{u}_{\text{a}} \right) - \left(\frac{U_{\text{a}}^r}{U_{\text{a}}^u}\right)^2 \frac{ \bar{u}_{\text{c}} + \bar{u}_{\text{b}} - 2 \bar{u}_{\text{a}} }{2} + \left[ \frac{\epsilon}{2} -  \epsilon \log \left(\frac{\bar{r}_{\text{a}}}{\bar{r}_{\text{p}}} \right) \right] \left( \bar{u}_{\text{b}} -  \bar{u}_{\text{a}} \right) \nonumber \\
 & + \sqrt{2 \epsilon } \, \text{erf}^{-1}\left\lbrace - \sqrt{\frac{2 \epsilon}{\pi}}  \exp\left[ - \frac{1}{2 \epsilon}  \left( \frac{U_{\text{a}}^r}{ U_{\text{a}}^u} \right)^2  \right]   \frac{\bar{u}_{\text{b}}-\bar{u}_{\text{a}}}{\bar{r}_{\text{a}}} + \text{erf}\left( \sqrt{\frac{1}{2 \epsilon} }  \frac{U_{\text{a}}^r}{U_{\text{a}}^u}\right)\right\rbrace \nonumber \\
 & \times \exp \left[ \frac{1}{2 \epsilon  } \left( \frac{U_{\text{a}}^r}{U_{\text{a}}^u} \right)^2 - \left( \text{erf}^{-1}\left\lbrace - \sqrt{\frac{2 \epsilon}{\pi}}  \exp\left[ - \frac{1}{2 \epsilon}  \left( \frac{U_{\text{a}}^r}{ U_{\text{a}}^u} \right)^2  \right]   \frac{\bar{u}_{\text{b}}-\bar{u}_{\text{a}}}{\bar{r}_{\text{a}}} + \text{erf}\left( \sqrt{\frac{1}{2 \epsilon} }  \frac{U_{\text{a}}^r}{U_{\text{a}}^u}\right)\right\rbrace \right)^2 \right] \bar{r}_{\text{a}} \nonumber \\
& + \epsilon  \left( \text{erf}^{-1}\left\lbrace - \sqrt{\frac{2 \epsilon}{\pi}}  \exp\left[ - \frac{1}{2 \epsilon}  \left( \frac{U_{\text{a}}^r}{ U_{\text{a}}^u} \right)^2  \right]   \frac{\bar{u}_{\text{b}}-\bar{u}_{\text{a}}}{\bar{r}_{\text{a}}} + \text{erf}\left( \sqrt{\frac{1}{2 \epsilon} }  \frac{U_{\text{a}}^r}{U_{\text{a}}^u}\right)\right\rbrace \right)^2 \left(\bar{u}_{\text{c}}-\bar{u}_{\text{b}}\right), \end{aligned}  \nonumber \\
& \begin{aligned} \bar{r}_\text{c} = \, &  \exp \left[ \frac{1}{2 \epsilon  } \left( \frac{U_{\text{a}}^r}{U_{\text{a}}^u} \right)^2 - \left( \text{erf}^{-1}\left\lbrace - \sqrt{\frac{2 \epsilon}{\pi}}  \exp\left[ - \frac{1}{2 \epsilon}  \left( \frac{U_{\text{a}}^r}{ U_{\text{a}}^u} \right)^2  \right]   \frac{\bar{u}_{\text{b}}-\bar{u}_{\text{a}}}{\bar{r}_{\text{a}}} + \text{erf}\left( \sqrt{\frac{1}{2 \epsilon} }  \frac{U_{\text{a}}^r}{U_{\text{a}}^u}\right)\right\rbrace \right)^2 \right] \bar{r}_{\text{a}} \nonumber \\
& + \sqrt{2 \epsilon } \, \text{erf}^{-1}\left\lbrace - \sqrt{\frac{2 \epsilon}{\pi}}  \exp\left[ - \frac{1}{2 \epsilon}  \left( \frac{U_{\text{a}}^r}{ U_{\text{a}}^u} \right)^2  \right]   \frac{\bar{u}_{\text{b}}-\bar{u}_{\text{a}}}{\bar{r}_{\text{a}}} + \text{erf}\left( \sqrt{\frac{1}{2 \epsilon} }  \frac{U_{\text{a}}^r}{U_{\text{a}}^u}\right)\right\rbrace \left(\bar{u}_{\text{c}}-\bar{u}_{\text{b}}\right), \end{aligned}  \nonumber \\
& \begin{aligned} & U_{\text{c}}^u = U_{\text{a}}^u, && U_{\text{c}}^v = U_{\text{a}}^v -\frac{\left(U_{\text{a}}^r\right)^2}{2 U_{\text{a}}^u}+ \epsilon  \left( \text{erf}^{-1}\left\lbrace - \sqrt{\frac{2 \epsilon}{\pi}}  \exp\left[ - \frac{1}{2 \epsilon}  \left( \frac{U_{\text{a}}^r}{ U_{\text{a}}^u} \right)^2  \right]   \frac{\bar{u}_{\text{b}}-\bar{u}_{\text{a}}}{\bar{r}_{\text{a}}} + \text{erf}\left( \sqrt{\frac{1}{2 \epsilon} }  \frac{U_{\text{a}}^r}{U_{\text{a}}^u}\right)\right\rbrace \right)^2 U_{\text{a}}^u , \end{aligned}  \nonumber \\
& U_{\text{c}}^r = \sqrt{2 \epsilon } \, \text{erf}^{-1}\left\lbrace - \sqrt{\frac{2 \epsilon}{\pi}}  \exp\left[ - \frac{1}{2 \epsilon}  \left( \frac{U_{\text{a}}^r}{ U_{\text{a}}^u} \right)^2  \right]   \frac{\bar{u}_{\text{b}}-\bar{u}_{\text{a}}}{\bar{r}_{\text{a}}} + \text{erf}\left( \sqrt{\frac{1}{2 \epsilon} }  \frac{U_{\text{a}}^r}{U_{\text{a}}^u}\right)\right\rbrace U_{\text{a}}^u.
\end{align}

\section{Sequence of pulses}\label{Sequence_of_pulses}

In this section, we consider a sequence of pulses, each with a fixed duration $\tau$, produced at a repetition rate $f_\text{rep}$. The $j$-th pulse is confined within $u \in [-\bar{z}_0/\sqrt{2} + (j-1)c/\sqrt{2}f_{\text{rep}},-\bar{z}_0/\sqrt{2} + (j-1)c/\sqrt{2}f_{\text{rep}} + c \tau/\sqrt{2} ]$, where $\bar{z}_0$ denotes the position of the first pulse at time $t=0$. We study the motion of a particle initially at rest and subsequently influenced by the sequence of pulses. At $t=0$, the particle interacts with the gravitational field of the first pulse. It is assumed that, during the observation period, the particle neither enters the pulses nor crosses the center $r=0$. Its trajectory is determined by applying the results from Sec.\ \ref{Particle_GW_interaction} to each spacetime region defined by $u \in [- \bar{z}_0/\sqrt{2} + (j-1)c /\sqrt{2}f_{\text{rep}},- \bar{z}_0/\sqrt{2} + j c/\sqrt{2}f_{\text{rep}}  ]$, and by patching the solutions across the boundaries at $u =- \bar{z}_0/\sqrt{2} + j c/\sqrt{2}f_{\text{rep}}$.

The trajectory of the particle is described by using the sequences $(-\bar{z}_0/\sqrt{2} + j c/\sqrt{2}f_{\text{rep}}, \bar{v}_j, \bar{r}_j, 0)$ and $U_j = (U_j^u, U_j^v, U_j^r, 0)$, indicating the position and the four-velocity at the beginning of the passage of the $(j+1)$-th GW in terms of the Minkowski coordinates $(u,v,r,\varphi)$. The recursive formula defining $(\bar{v}_j, \bar{r}_j, U_j^u, U_j^v, U_j^r)$ can be obtained by replacing $\bar{u}_{\text{c}}\mapsto [-\bar{z}_0 + j c /f_{\text{rep}}]/\sqrt{2},\bar{u}_{\text{b}}\mapsto [-\bar{z}_0 + (j-1)c/f_{\text{rep}}+c \tau]/\sqrt{2} ,\bar{u}_{\text{a}}\mapsto [-\bar{z}_0 + (j-1) c / f_{\text{rep}}]/\sqrt{2}, \bar{v}_{\text{c}}\mapsto \bar{v}_j,\bar{r}_{\text{c}}\mapsto \bar{r}_j,U_{\text{c}}^u\mapsto U_j^u,U_{\text{c}}^v\mapsto U_j^v,U_{\text{c}}^r\mapsto U_j^r, \bar{v}_{\text{a}}\mapsto \bar{v}_{j-1},\bar{r}_{\text{a}}\mapsto \bar{r}_{j-1},U_{\text{a}}^u\mapsto U_{j-1}^u,U_{\text{a}}^v\mapsto U_{j-1}^v,U_{\text{a}}^r\mapsto U_{j-1}^r$ into Eq.\ (\ref{gamma_c}). The resulting equations explicitly read
\begin{align}\label{gamma_U_j}
 \bar{v}_j = \, & \bar{v}_{j-1}  - \frac{U_{j-1}^r}{U_{j-1}^u} \bar{r}_{j-1}  + \frac{1}{\sqrt{2}} \frac{c}{f_{\text{rep}}}\frac{U_{j-1}^v}{U_{j-1}^u}  - \frac{1}{2\sqrt{2}} \left( \frac{c}{f_{\text{rep}}} + c \tau \right) \left(\frac{U_{j-1}^r}{U_{j-1}^u}\right)^2 + \frac{\epsilon c \tau}{\sqrt{2}}  \left[ \frac{1}{2} -   \log \left(\frac{\bar{r}_{j-1}}{\bar{r}_{\text{p}}} \right)  \right]  \nonumber \\
 & + \sqrt{2 \epsilon } \, \text{erf}^{-1}\left\lbrace - \sqrt{\frac{\epsilon}{\pi}}  \frac{c \tau}{\bar{r}_{j-1}}  \exp\left[ - \frac{1}{2 \epsilon}  \left( \frac{U_{j-1}^r}{ U_{j-1}^u} \right)^2  \right]  + \text{erf}\left( \sqrt{\frac{1}{2 \epsilon} }  \frac{U_{j-1}^r}{U_{j-1}^u}\right)\right\rbrace \nonumber \\
 & \times \exp \left[ \frac{1}{2 \epsilon  } \left( \frac{U_{j-1}^r}{U_{j-1}^u} \right)^2 - \left( \text{erf}^{-1}\left\lbrace - \sqrt{\frac{\epsilon}{\pi}}  \frac{c \tau}{\bar{r}_{j-1}}  \exp\left[ - \frac{1}{2 \epsilon}  \left( \frac{U_{j-1}^r}{ U_{j-1}^u} \right)^2  \right]  + \text{erf}\left( \sqrt{\frac{1}{2 \epsilon} }  \frac{U_{j-1}^r}{U_{j-1}^u}\right)\right\rbrace \right)^2 \right] \bar{r}_{j-1} \nonumber \\
& +  \frac{\epsilon}{\sqrt{2}}  \left( \text{erf}^{-1}\left\lbrace - \sqrt{\frac{\epsilon}{\pi}}  \frac{c \tau}{\bar{r}_{j-1}}  \exp\left[ - \frac{1}{2 \epsilon}  \left( \frac{U_{j-1}^r}{ U_{j-1}^u} \right)^2  \right]  + \text{erf}\left( \sqrt{\frac{1}{2 \epsilon} }  \frac{U_{j-1}^r}{U_{j-1}^u}\right)\right\rbrace \right)^2 \left( \frac{c}{f_{\text{rep}}} - c \tau \right),   \nonumber \\
 \bar{r}_j = \, &  \exp \left[ \frac{1}{2 \epsilon  } \left( \frac{U_{j-1}^r}{U_{j-1}^u} \right)^2 - \left( \text{erf}^{-1}\left\lbrace - \sqrt{\frac{\epsilon}{\pi}}  \frac{c \tau}{\bar{r}_{j-1}}  \exp\left[ - \frac{1}{2 \epsilon}  \left( \frac{U_{j-1}^r}{ U_{j-1}^u} \right)^2  \right]  + \text{erf}\left( \sqrt{\frac{1}{2 \epsilon} }  \frac{U_{j-1}^r}{U_{j-1}^u}\right)\right\rbrace \right)^2 \right] \bar{r}_{j-1} \nonumber \\
& + \sqrt{\epsilon } \, \text{erf}^{-1}\left\lbrace - \sqrt{\frac{\epsilon}{\pi}}  \frac{c \tau}{\bar{r}_{j-1}}  \exp\left[ - \frac{1}{2 \epsilon}  \left( \frac{U_{j-1}^r}{ U_{j-1}^u} \right)^2  \right]  + \text{erf}\left( \sqrt{\frac{1}{2 \epsilon} }  \frac{U_{j-1}^r}{U_{j-1}^u}\right)\right\rbrace \left( \frac{c}{f_{\text{rep}}} - c \tau \right),  \nonumber \\
  U_j^u = \, & \begin{aligned}  & U_{j-1}^u, & U_j^v = U_{j-1}^v -\frac{\left(U_{j-1}^r\right)^2}{2 U_{j-1}^u}+ \epsilon   \left( \text{erf}^{-1}\left\lbrace - \sqrt{\frac{\epsilon}{\pi}}  \frac{c \tau}{\bar{r}_{j-1}}  \exp\left[ - \frac{1}{2 \epsilon}  \left( \frac{U_{j-1}^r}{ U_{j-1}^u} \right)^2  \right]  + \text{erf}\left( \sqrt{\frac{1}{2 \epsilon} }  \frac{U_{j-1}^r}{U_{j-1}^u}\right)\right\rbrace \right)^2 U_{j-1}^u, \end{aligned} \nonumber \\
 U_j^r =\, & \sqrt{2 \epsilon }\, \text{erf}^{-1}\left\lbrace - \sqrt{\frac{\epsilon}{\pi}}  \frac{c \tau}{\bar{r}_{j-1}}  \exp\left[ - \frac{1}{2 \epsilon}  \left( \frac{U_{j-1}^r}{ U_{j-1}^u} \right)^2  \right]  + \text{erf}\left( \sqrt{\frac{1}{2 \epsilon} }  \frac{U_{j-1}^r}{U_{j-1}^u}\right)\right\rbrace U_{j-1}^u  .
\end{align}
The initial conditions are imposed by assuming $\bar{v}_0 = \bar{z}_0/\sqrt{2}$, $ U_0^u = 1/\sqrt{2}$, $ U_0^v = 1/\sqrt{2}$, $ U_0^r = 0$, and by choosing the initial position through $\bar{z}_0$ and $\bar{r}_0$.

Equation (\ref{gamma_U_j}) cannot be solved explicitly, unless we consider the weak gravity approximation ($\epsilon \ll 1$). To obtain such a result, we expand each variable as
\begin{align}\label{expansion_epsilon_j}
& \bar{v}_{j}=\bar{v}_{j,0}+ \epsilon  \bar{v}_{j,1}+O\left(\epsilon^2\right), && \bar{r}_{j}= \bar{r}_{j,0}+\epsilon  \bar{r}_{j,1}+O\left(\epsilon^2\right), && U_{j}^u= U_{j,0}^u+\epsilon  U_{j,1}^u+O\left(\epsilon^2\right), \nonumber \\
& U_{j}^v= U_{j,0}^v + \epsilon U_{j,1}^v+ \epsilon ^2 U_{j,2}^v + O\left(\epsilon^3\right), && U_{j}^r= U_{j,0}^r + \epsilon  U_{j,1}^r+O\left(\epsilon^2\right).
\end{align} 
The initial conditions are
\begin{align}\label{expansion_epsilon_0}
& \bar{v}_{0,0} = \frac{\bar{z}_0}{\sqrt{2}}, &&  \bar{v}_{0,1} = 0, && \bar{r}_{0,0} = \bar{r}_{0}, && \bar{r}_{0,1}=0, && U_{0,0}^u = \frac{1}{\sqrt{2}}, && U_{0,1}^u = 0, \nonumber \\
&  U_{0,0}^v = \frac{1}{\sqrt{2}}, && U_{0,1}^v=0, && U_{0,2}^v=0, && U_{0,0}^r=0, &&  U_{0,1}^r=0.
\end{align}

By imposing the ansatz $U_{j,0}^r=0$ and by plugging Eq.\ (\ref{expansion_epsilon_j}) into Eq.\ (\ref{gamma_U_j}), we obtain the recursive formula
\begin{align}\label{gamma_U_j_epsilon}
& \begin{aligned} & \bar{v}_{j,0} = \bar{v}_{j-1,0} + \frac{1}{\sqrt{2}}\frac{U_{j-1,0}^v}{U_{j-1,0}^u} \frac{c}{f_{\text{rep}}} , & \bar{v}_{j,1} = \bar{v}_{j-1,1} - \frac{1}{\sqrt{2}} \left[ \frac{1}{2}  + \log \left(\frac{\bar{r}_{j-1,0}}{\bar{r}_{\text{p}}}\right) \right] c \tau + \frac{1}{\sqrt{2}} \left[ \frac{U_{j-1,1}^v}{U_{j-1,0}^u} - \frac{U_{j-1,1}^u U_{j-1,0}^v}{\left(U_{j-1,0}^u\right)^2} \right] \frac{c}{f_{\text{rep}} } , \end{aligned} \nonumber \\ 
& \begin{aligned} & \bar{r}_{j,0} = \bar{r}_{j-1,0}, && \bar{r}_{j,1}= \bar{r}_{j-1,1} + \frac{1}{\sqrt{2}} \frac{c}{f_{\text{rep}}} \frac{U_{j-1,1}^r}{U_{j-1,0}^u}  + \frac{1}{2} \left( \frac{ c^2 \tau ^2 }{2} - \frac{c^2 \tau}{f_{\text{rep}}} \right) \frac{1}{\bar{r}_{j-1,0}}, && U_{j,0}^u = U_{j-1,0}^u,  && U_{j,1}^u = U_{j-1,1}^u,  \end{aligned} \nonumber \\ 
& \begin{aligned} & U_{j,0}^v = U_{j-1,0}^v,  && U_{j,1}^v=U_{j-1,1}^v, && U_{j,2}^v=U_{j-1,2}^v +  \left( \frac{U_{j-1,0}^u}{4} \frac{c \tau}{\bar{r}_{j-1,0}}  -\frac{U_{j-1,1}^r}{\sqrt{2}}  \right) \frac{c \tau }{ \bar{r}_{j-1,0}} ,   \end{aligned} \nonumber \\ 
& \begin{aligned} & U_{j,0}^r=0,  && U_{j,1}^r= U_{j-1,1}^r- \frac{U_{j-1,0}^u}{\sqrt{2}} \frac{c \tau}{\bar{r}_{j-1,0}} . \end{aligned} 
\end{align}
The solution of Eq.\ (\ref{gamma_U_j_epsilon}) with initial conditions given by Eq.\ (\ref{expansion_epsilon_0}) is
\begin{align}\label{gamma_U_j_epsilon_solution}
& \begin{aligned} & \bar{v}_{j,0} = \frac{\bar{z}_{0}}{\sqrt{2}} + \frac{j}{\sqrt{2}} \frac{c}{f_{\text{rep}}}, && \bar{v}_{j,1} = -\frac{j}{\sqrt{2}}  \left[\frac{1}{2}+ \log \left(\frac{\bar{r}_0}{\bar{r}_{\text{p}}}\right)\right] c  \tau , && \bar{r}_{j,0} = \bar{r}_{0}, && \bar{r}_{j,1}= \frac{j}{4} \frac{c^2 \tau^2}{\bar{r}_0 } - \frac{j(j+1)}{4}  \frac{c^2 \tau}{\bar{r}_0 f_{\text{rep}}}, \end{aligned} \nonumber \\ 
& \begin{aligned} & U_{j,0}^u = \frac{1}{\sqrt{2}},  && U_{j,1}^u = 0,  && U_{j,0}^v = \frac{1}{\sqrt{2}},  && U_{j,1}^v= 0,  && U_{j,2}^v=  \frac{j^2}{4 \sqrt{2} }  \frac{c^2 \tau^2}{\bar{r}_0^2}, && U_{j,0}^r=0,  && U_{j,1}^r= -\frac{j }{2} \frac{c \tau}{\bar{r}_0}. \end{aligned} 
\end{align}
This proves Eqs.\ (11) and (12) of the main paper.

\bibliographystyle{apsrev4-2}
\bibliography{bibliography}

\begin{figure}[h]
\includegraphics[]{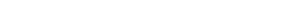}
\end{figure}